\documentclass[lettersize,journal]{IEEEtran}
\usepackage{array}
\usepackage{soul}
\usepackage[caption=false,font=normalsize,labelfont=sf,textfont=sf]{subfig}
\usepackage{algorithm,algorithmicx,algpseudocode}
\usepackage{enumitem}
\usepackage{listings}
\usepackage{graphicx}
\usepackage{tabularx}
\usepackage[labelfont={bf}]{caption}
\usepackage{lineno}
\usepackage{amsmath}
\usepackage{amsthm}
\usepackage{mathtools}
\usepackage{amsfonts}
\usepackage{amssymb}
\usepackage{xcolor}
\usepackage{orcidlink}

\newcommand\blfootnote[1]{%
  \begingroup
  \renewcommand\thefootnote{}\footnote{#1}%
  \addtocounter{footnote}{-1}%
  \endgroup
}

\newcommand{\dm}[1] {}
\newcommand{\cj}[1] {}

\newtheorem{Claim}{Claim}

\makeatletter
\newcommand{\ostar}{\mathbin{\mathpalette\make@circled\star}}
\newcommand{\make@circled}[2]{%
	\ooalign{$\m@th#1\smallbigcirc{#1}$\cr\hidewidth$\m@th#1#2$\hidewidth\cr}%
}
\newcommand{\smallbigcirc}[1]{%
	\vcenter{\hbox{\scalebox{0.77778}{$\m@th#1\bigcirc$}}}%
}
\makeatother
\usepackage[cmintegrals]{newtxmath}

\def\BibTeX{{\rm B\kern-.05em{\sc i\kern-.025em b}\kern-.08em
    T\kern-.1667em\lower.7ex\hbox{E}\kern-.125emX}}
\usepackage{balance}
\begin{document}
\title{On the Coverage Required for Diploid Genome Assembly}

\author{
	\IEEEauthorblockN{Daanish Mahajan}, \IEEEauthorblockN{Chirag Jain}, \IEEEauthorblockN{Navin Kashyap}

\thanks{Daanish Mahajan is with the Department of Interdisciplinary Mathematical Sciences, Indian Institute of Science, Bangalore 560012 India (e-mail: \href{daanishm@iisc.ac.in}{daanishm@iisc.ac.in})}

\thanks{Chirag Jain is with the Department of Computational and Data Sciences, Indian Institute of Science, Bangalore 560012 India (e-mail: \href{chirag@iisc.ac.in}{chirag@iisc.ac.in})}

\thanks{Navin Kashyap is with the Department of Electrical Communication Engineering, Indian Institute of Science, Bangalore 560012 India (e-mail: \href{nkashyap@iisc.ac.in}{nkashyap@iisc.ac.in})}
}

\markboth{Journal of \LaTeX\ Class Files,~Vol.~18, No.~9, September~2020}%
{How to Use the IEEEtran \LaTeX \ Templates}

\maketitle

\begin{abstract}
  The repeat content and heterozygosity rate of a target genome are important factors in determining the feasibility of achieving a complete telomere-to-telomere assembly. 
  %Repeat content and heterozygosity rate of the target genome are crucial factors in determining the ability to generate a complete telomere-to-telomere assembly. 
  The mathematical relationship between the required coverage and read length for the purpose of unique reconstruction remains unexplored for diploid genomes. We investigate the information-theoretic conditions that the given set of sequencing reads must satisfy to achieve the complete reconstruction of the true sequence of a diploid genome. We also analyze the standard greedy and de-Bruijn graph-based assembly algorithms. 
  %and compare the coverage depth and read length requirements with the information-theoretic lower bound. 
  %Our results show that the gap between the two is considerable because both algorithms require the double repeats in the genome to be bridged.
  Our results show that the coverage and read length requirements of the assembly algorithms are considerably higher than the lower bound because both algorithms require the double repeats in the genome to be bridged.
  Finally, we derive the necessary conditions for the overlap graph-based assembly paradigm.\blfootnote{A preliminary version of this work appeared in ISIT'24 \cite{mahajan2024coverage}}
  %\CJ{This abstract was okay for ISIT but needs to be revised for comp-bio journals. It needs to be a bit more verbose. Add motivation in the beginning to make the reader understand why your work is important.} \CJ{You should add a footnote on the first page saying that a preliminary version of this work appeared in ISIT 2024 [cite]}
\end{abstract}


\section{Introduction}
The problem of reconstructing genomes, i.e., \emph{de novo} genome assembly, is a fundamental problem in computational biology. DNA sequencing instruments cannot produce an individual's entire DNA sequence (genome) but produce many short strings (reads) sampled from random unknown genome locations. The \emph{de novo} genome assembly problem seeks to reconstruct the original genome using the set of reads as input. The length and accuracy of reads depend on the choice of sequencing technology. Recent long-read sequencing instruments can produce reads of length above $10$~kbp with a sequencing error rate below one percent \cite{logsdon2020long, marx2023method}. 
%If reads are assumed to have a fixed length $L$, a lower bound on the count of reads needed to cover the entire genome sequence can be obtained using Lander-Waterman statistics \cite{lander1988genomic}. 
If a genome sequence of length $G$ is to be covered entirely with probability $1 - \epsilon$ using reads of length $L$, then a lower bound on the count of reads is available from the Lander-Waterman statistics \cite{lander1988genomic,motahari2013information}, which is approximately $\frac{G}{L}\ln{\frac{G}{L\epsilon}}$. In the genome assembly problem, these reads are to be `stitched together' without prior knowledge about the initial order in which they were present.\vspace*{0.1cm}

Early works have formulated genome assembly as either (i) finding the shortest common superstring of the set of reads or (ii) finding a Hamiltonian path of a desired length in a read-overlap graph, both of which are known to be NP-hard \cite{shomorony2016information}. These theoretical models often fail to work on real data because of the size of the data and errors in the genome sequencing process \cite{medvedev2019modeling}. Also, if repeats in a genome are longer than the length of reads, the parsimony-based approaches can affect the count of these repeats in the reconstructed sequence compared to the true sequence. 
Some practical efforts to solve the assembly problem include building a de-Bruijn graph and finding an Eulerian Superpath of the set of reads in the obtained graph \cite{pevzner2001eulerian}. The other approach includes building a read-overlap graph \cite{myers2005fragment}. 
%The final goal here is to convert the problem into a flow problem, ensuring that the total length of the genome reconstructed is minimized.
%\CJ{Before mentioning assemblers Hifiasm/Verkko, maybe you could briefly introduce greedy algorithm / de Bruijn / overlap graphs and their connection to SOTA. This may help justify why you studied these. Please also cite the original papers proposing these algorithms.} 
State-of-the-art assemblers, e.g., Hifiasm~\cite{cheng2023scalable}, Verkko~\cite{rautiainen2023telomere}, though built on top of the overlap graph and de Bruijn graph frameworks respectively, combine several other heuristics and multiple sequencing technologies to assemble reads into significantly longer substrings (contigs) of the genome.  
%These contigs may need further processing for scaffolding and gap filling \cite{li2023genome}. 
All this machinery is required because of the fundamental barrier of long repeats in a genome that cannot be bridged by a single read. This limitation can make the underlying genome's unique reconstruction impossible.

This is the basis of the information-optimal genome assembly problem, which asks for the minimum amount of information (read length and coverage depth) required under which the genome can be reconstructed successfully, i.e., with information less than this, no algorithm can guarantee reconstruction of the unknown genome sequence. Hence, these necessary conditions impose a fundamental limit on the performance of any assembly algorithm. Once we have the information-theoretic necessary conditions, the next task is to identify efficient algorithms and analyse the amount of information required (necessary and sufficient conditions) by them to reconstruct the ground truth genome.  Motahari \emph{et al.} \cite{motahari2013information} gave information-theoretic necessary conditions for genome assembly under i.i.d. and Markov model for DNA sequence. The follow-up work in \cite{shomorony2016information,bresler2013optimal}, and \cite{hui2016overlap} provided the necessary conditions for arbitrary repeat statistics and gave sufficient conditions for the known assembly algorithms that are based on greedy, de Bruijn graph and overlap graph-based approaches. % \CJ{Somewhere in the above paragraph, would it help to say what do you mean by `sufficient conditions', similar to how you have defined `necessary conditions' above.}
%They derived sufficient conditions for greedy and $k$-mer based algorithms.
% They classified the repeats into two broad categories: 1) Triple and 2) Interleaved repeats, and necessary conditions ask for the bridging of these repeats. The sufficient condition for the greedy algorithm is to have all the repeats bridged and for the de Bruijn graph-based algorithm is to have the length of $k-$mer strictly greater than the length of the triple and interleaved repeats, and the de Bruijn graph formed to be strongly connected.
% The sufficient condition for the de Bruijn graph-based algorithm is to have the length of $k-$mer strictly greater than the length of the triple and interleaved repeats, and for the overlap-based algorithm is to have triple repeats to be all-bridged and interleaved repeats to be bridged. Also, the final de Bruijn graph and the overlap graph formed must be strongly connected. A crucial step while dealing with overlap graphs is to convert a node-covering walk (that does not have a poly-time algorithm) to an edge-covering walk (that has a poly-time algorithm) by pruning the overlap graph to remove the redundant edges. The main advantage that one gets by the assumption of triple repeats being all-bridged is that one can bound the in-degree and out-degree of vertices to at most $2$. It can be shown that the degrees cannot be bounded once this restriction is removed.

The above works considered the \emph{haploid} genome assembly problem. A haploid genome includes a single copy of each chromosome. There is limited information-theoretic research for \emph{diploid} genomes, i.e., when the unknown genome comprises two near-identical copies of each chromosome inherited from both parents. 
%\CJ{``There is limited work for diploid genomes, i.e.,..." Should this be ``There is limited information-theoretic research for diploid genomes, i.e.,..."}. 
For example, human genomes are diploid. The set of chromosomes inherited from a single parent is called a \emph{haplotype}. The combined length of human chromosomes is about 3 billion nucleotides in each haplotype. The two haplotypes in a human genome are about $99.9\%$ identical. These haplotypes differ due to mutations in the form of substitutions, insertions, deletions, etc. The loci at which the two haplotypes differ are referred to as \emph{heterozygous} loci. Identifying heterozygous loci imposes an additional challenge for any diploid genome assembler because a read does not contain information about the haplotype from which they originate. In case the read length is smaller than the maximum gap between two adjacent heterozygous loci, one obvious error an algorithm can make is to switch between paternal and maternal haplotypes during genome reconstruction. Separating the two haplotypes (also known as \emph{phasing}) is usually a post-processing step that requires long-range Hi-C data or sequencing data of the parents \cite{kronenberg2021extended}. Si \emph{et al.} \cite{si2017information} addressed a related problem of reference-based diploid genome assembly where a reference genome is used during assembly.

\begin{figure}[htpb]
    \begin{center}
        \includegraphics[scale=0.32]{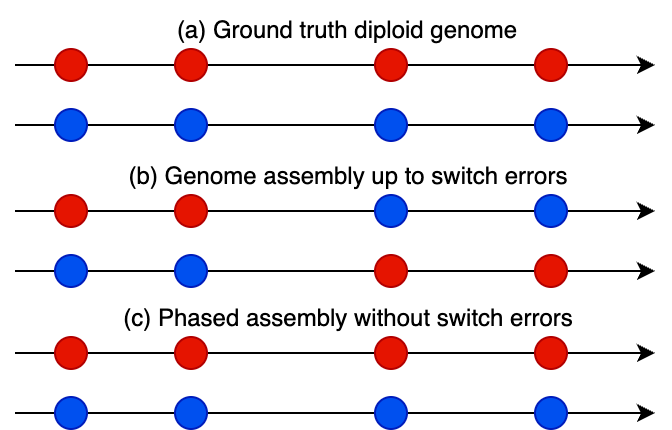}
        \caption{Illustration of genome assembly with and without switch errors. The red and blue circles are used to represent nucleotides at heterozygous loci.}
        \label{assembly-types}
    \end{center}
\end{figure}

%To the best of our knowledge, the only work done under a diploid setting is by Si \emph{et al.} \cite{si2017information}. This study finds the number of mate-paired reads required to reconstruct haplotypes under both error-free and i.i.d. error models with the assumption that each read gives information about the nucleotides present on at least two heterozygous loci. 
%They derived the number of mate-paired reads required to reconstruct haplotypes under both error-free and i.i.d. error models with the assumption that each read gives information about the nucleotides present on at least two heterozygous loci. 
%\CJ{Si \emph{et al.} do not address the \emph{de novo} assembly problem. Their work assumes that reads have been aligned to a reference genome. They are solving the haplotype phasing problem, i.e., their goal is to identify the variants that are co-located on the same haplotype. This is a related but different problem. Please revise this sentence. How about we say :  Si \emph{et al.} \cite{si2017information} considered a related problem of reference-based assembly of a diploid genome where a reference genome is used to guide the genome reconstruction. They derived.... \quad In this paper, we focus on the \emph{de novo} diploid genome assembly problem..}.
In this paper, we focus on \emph{de novo} diploid genome assembly problem from an information-theoretic perspective. We derive (a) information-theoretic necessary conditions, (b) necessary and sufficient conditions for the greedy and de Bruijn graph-based assembly algorithms, and (c) necessary conditions for the overlap graph-based algorithm such that the assembled genome matches the ground truth up to switch errors. Figure \ref{assembly-types}(b) gives an illustration of an assembled diploid genome that matches the ground truth up to switch errors. An assembly with switch errors is also called as a partially phased assembly in bioinformatics literature \cite{heng2021}.
%Our information-theoretic necessary conditions are almost similar to the known conditions for the haploid genome case \cite{bresler2013optimal}. 

We prove that all three algorithms (i.e., greedy, de Bruijn graph-based, and overlap graph-based algorithms) require bridging of \emph{double repeats} to guarantee the reconstruction of the diploid genome up to switch errors. Our theoretical results imply that the information required by the assembly algorithms in terms of coverage depth and read length is significantly more for reconstructing diploid genomes than haploid genomes. We empirically evaluate the algorithms by using repeat statistics of a publicly available assembled sequence of human chromosome 19. This experiment sheds light on the information requirements of the assembly algorithms for an unassembled genome having similar statistics.

%\DM{We verify these results empirically in our experiment (Section \ref{experiments}) with human T2T chromosome $19$.} %where \emph{interleaved} and \emph{triple repeats} are a bottleneck . 

% Commenting this paragraph to save space [CJ]
%The remainder of the paper is organized as follows: Section \ref{problem-statement} introduces notations, definitions and describe the problem statement. Section \ref{it-necessary-conditions} presents the information-theoretic necessary conditions. Section \ref{algorithms} presents the algorithm specific necessary and sufficient conditions. Section \ref{experiments} summarizes the experimental results. Finally, we conclude in Section \ref{conclusions}.

\section{Preliminaries}
\subsection{Notations and Definitions}
\label{notations}
\begin{figure}[htpb]
    \begin{center}
        \includegraphics[scale=0.25]{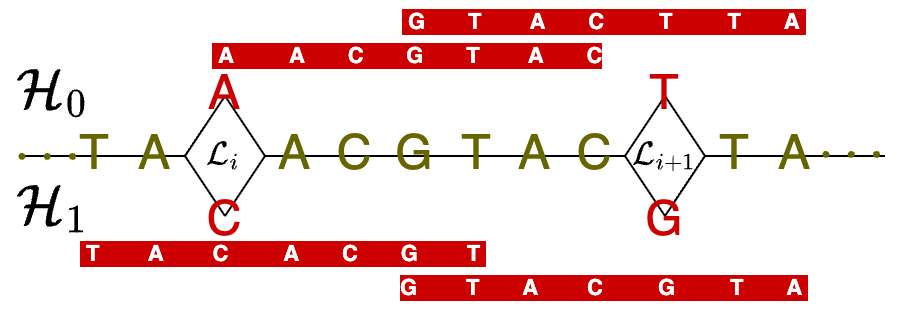}
        \caption{An example illustrating four reads sampled from a diploid genome $(\mathcal{H}_0, \mathcal{H}_1)$. $\mathcal{L}_i$ and $\mathcal{L}_{i+1}$ are heterozygous loci.}
        \label{diploid-enumeration}
    \end{center}
\end{figure}

We assume that the unknown diploid genome comprises two circular strings. %, i.e., two near-identical copies of a single-stranded chromosome. 
We enumerate the unknown pair of haplotype sequences as $(\mathcal{H}_0, \mathcal{H}_1)$ of length $\lvert \mathcal{H} \rvert$ each. Suppose the two haplotypes differ at exactly $n$ heterozygous loci (where $n$ is also unknown). We assume that the variation between the haplotypes is only due to substitution events. We enumerate the heterozygous loci by $\mathcal{L}_1, \mathcal{L}_2, \ldots, \mathcal{L}_n$ in a clockwise manner starting from an arbitrary locus. 
Accordingly, $\mathcal{H}_0[i] = \mathcal{H}_1[i]$ for all $i \in \{0, \ldots, \lvert \mathcal{H} \rvert-1\} \setminus \{\mathcal{L}_1, \ldots, \mathcal{L}_n\}$. 
Similarly, $\mathcal{H}_0[i] \neq \mathcal{H}_1[i]$ for all $i \in \{\mathcal{L}_1, \ldots, \mathcal{L}_n\}$. Since the genome is circular, we define $\mathcal{L}_0 := \mathcal{L}_n$ and $\mathcal{L}_{n + 1} := \mathcal{L}_1$. 
%We have an unknown diploid single-stranded circular genome $\mathcal{G}$ enumerated by a pair of haplotype sequences $(\mathcal{H}_0, \mathcal{H}_1)$ of length $\lvert \mathcal{H} \rvert$ each. Therefore, both haplotypes are assumed to be circular strings. These haplotypes vary at exactly $n$ positions (where $n$ is unknown), known as heterozygous loci. We enumerate them by $\mathcal{L}_1, \mathcal{L}_2, \ldots, \mathcal{L}_n$ in a clockwise manner starting from an arbitrary locus. Therefore, $\mathcal{L}_{n + 1} = \mathcal{L}_1$ and $\mathcal{L}_0 = \mathcal{L}_n$. 
 %\hl{Similar to} \cite{shomorony2016information}, 
 We make simplication assumptions that all input reads in a given set, say $\mathcal{R}$, are error-free, single-stranded, and have fixed length $L$. Each read is sampled uniformly and independently from one of the two haplotypes (Figure \ref{diploid-enumeration}). The average number of times a nucleotide in the genome is sampled during a sequencing experiment is referred to as \emph{coverage depth}. It can be calculated as $\frac{\lvert\mathcal{R}\rvert L}{2\lvert \mathcal{H} \rvert}$.  

Assuming $0$-based indexing, $s[x:y]$ denotes a substring of $s$ from index $x$ to index $y - 1$, both inclusive. Likewise, $s[x: ]$ denotes a suffix of $s$ starting from index $x$ and $s[:y] (y > 0)$ denotes a prefix of $s$ ending at index $y - 1$. 
$\textsc{overlap}(x, y)$ returns the maximum length of a proper suffix of read $x$ that is also a proper prefix of read $y$. $\textsc{union}(x, y)$ returns the concatenation of strings $x$ and $y$ after removing the common overlap substring from prefix of $y$, i.e., $\textsc{union}(x, y) = x + y[\textsc{overlap}(x, y):]$, where `+' is the concatenation operator. 
% of the type $\mathcal{H}_i[s : e]$ (where $s \geq e$) denotes $\mathcal{H}_i[s : \lvert \mathcal{H} \rvert] + \mathcal{H}_i[0 : e]$.

\begin{figure}[htbp]
    \centering
    \includegraphics[scale=0.19]{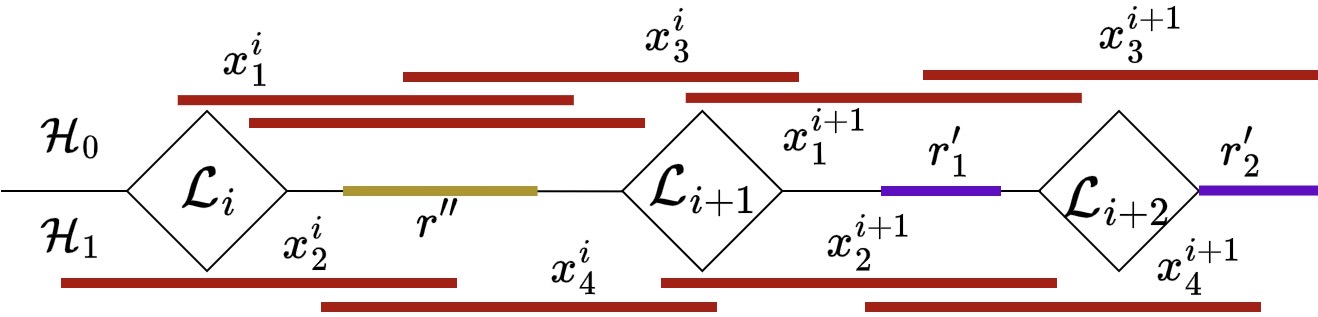}
    \caption{Reads $x_1^i$ and $x_2^i$ cover the heterozygous locus $\mathcal{L}_i$ and extend the maximum towards $\mathcal{L}_{i+1}$. Similarly, reads $x_3^i$ and $x_4^i$ cover the heterozygous locus $\mathcal{L}_{i+1}$ and extend the maximum towards $\mathcal{L}_{i}$.  Substring $r''$ (shown in yellow) is double-bridged by reads $x_1^i$ and $x_4^i$. Double repeat $(r_1', r_2')$ (shown in purple) is well-bridged by reads $x_1^{i + 1}$ and $x_2^{i + 1}$.}
    \label{diploid-well-bridging}
\end{figure}

For every ordered pair $(\mathcal{L}_i, \mathcal{L}_{i + 1})$ of two adjacent heterozygous loci, enumerate $x_1^i$ (resp.\ $x_2^i$) as the read covering $\mathcal{L}_i$ in haplotype $\mathcal{H}_0$ (resp.\ $\mathcal{H}_1$) with its right end-point being the farthest in the clockwise direction among all such reads. See Figure \ref{diploid-well-bridging} for an example. Similarly, enumerate $x_3^i$ (resp.\ $x_4^i$) as the read covering $\mathcal{L}_{i + 1}$ in haplotype $\mathcal{H}_0$ (resp.\ $\mathcal{H}_1$) with its left end-point being the farthest in the anti-clockwise direction among all such reads.

% \begin{figure}[hbt!]
%     % \centering
%     \begin{center}
%         \includegraphics[scale=0.2]{images/diploid-enumeration.png}
%         \caption{Figure enumerates the reads covering the heterozygous locus that extends the maximum towards the opposite locus.}
%         \label{diploid-enumeration}
%     \end{center}
% \end{figure}
% consisting of two haplotypes $\mathcal{H}_0$ and $\mathcal{H}_1$

%\subsection{Definitions}
%We use $[x]$ to denote the set $\{1, 2, \ldots, x\}, x\in \mathbb{N}$. 
Since the genome is circular, indexing is considered modulo $\lvert \mathcal{H} \rvert$. Next, we define the different types of repeats that will be used in our analysis.
Let $s_{i, j}^l$ denote a substring of length $l$ ($l \in \{1,2, \ldots, \lvert \mathcal{H} \rvert\}$) in haplotype $\mathcal{H}_i$ ($i \in \{0, 1\}$) starting at position $j$ ($j \in \{0, 1, 2, \ldots, \lvert \mathcal{H} \rvert - 1\}$). 
We refer to a pair of strings $(s_{i_1, j_1}^l, s_{i_2, j_2}^l)$ as a \emph{double repeat} if the following conditions hold: (a) $s_{i_1, j_1}^l = s_{i_2, j_2}^l$, (b) $j_1 < j_2$, and (c) $\mathcal{H}_{i_1}[j_1 - 1] \neq \mathcal{H}_{i_2}[j_2 - 1]$ and $\mathcal{H}_{i_1}[j_1 + l] \neq \mathcal{H}_{i_2}[j_2 + l]$. The definition requires that the match is maximal. The subtype of double repeat with $i_1 = i_2$ is called intra-double repeat, and the one with $i_1 \neq i_2$ is called inter-double repeat. 
A 3-tuple of strings $(s_{i_1, j_1}^l, s_{i_2, j_2}^l, s_{i_3, j_3}^l)$ is known as a \emph{triple repeat} if the following conditions hold: (a) $s_{i_1, j_1}^l = s_{i_2, j_2}^l = s_{i_3, j_3}^l$ and $j_1 < j_2 < j_3$, (b) Neither $\mathcal{H}_{i_1}[j_1 - 1] = \mathcal{H}_{i_2}[j_2 - 1] = \mathcal{H}_{i_3}[j_3 - 1]$ nor $\mathcal{H}_{i_1}[j_1 + l] = \mathcal{H}_{i_2}[j_2 + l] = \mathcal{H}_{i_3}[j_3 + l]$. The subclass of triple repeat with $i_1 = i_2 = i_3$ is called intra-triple repeat.
Two pairs of strings $(s_{i_1, j_1}^{l_1}, s_{i_2, j_2}^{l_1})$ and $(s_{i_3, j_3}^{l_2}, s_{i_4, j_4}^{l_2})$ are referred to as an \emph{interleaved repeat} if they are individually double repeats such that $j_1 < j_3 < j_2 < j_4$ and $s_{i_1, j_1}^{l_1} \neq s_{i_3, j_3}^{l_2}$. The subclass of interleaved repeat with $i_1 = i_2 = i_3 = i_4$ is called intra-interleaved repeat.

A substring $s_{i, j}^l(l \leq L)$ is said to be \emph{covered} by a read $r$ if $\exists k \in [j + l - L, j]$ such that $s_{i, k}^L = r$. For example, in Figure  \ref{diploid-well-bridging}, substring $r_2'$ is covered by read $x_3^{i + 1}$. A set of reads is said to cover the diploid genome if, for both haplotypes individually, every pair of adjacent nucleotides is covered by at least one read. 
%\hl{Coverage depth ($c = \frac{NL}{G}$) is defined as the average number of reads covering a nucleotide in a genome. Here $N$ denotes the number of reads required and depends on the algorithm used.}
%\emph{Coverage depth} is the average number of reads covering a nucleotide in a genome.
A substring in the genome is said to be \emph{bridged} by a read if it covers at least one base just before and one just after that substring. For example, in Figure \ref{diploid-well-bridging}, substring $r''$ is bridged by read $x_1^i$. A double repeat or a triple repeat is said to be bridged if at least one copy of the repeat is bridged. An interleaved repeat is said to be bridged if at least one of the double repeat is bridged.
% , all-bridged if all copies are bridged, and all-unbridged if none are bridged. 

A substring in the genome is said to be \emph{double-bridged} by reads $r_1, r_2$ if the substring is bridged by both $r_1, r_2$. 
A double repeat is said to be \emph{well-bridged} if at least one of the following is satisfied: (a) At least one repeat copy covers a heterozygous locus and is bridged, (b) At least one repeat copy does not cover a heterozygous locus and is double-bridged by two reads which cover a common heterozygous locus on opposite haplotypes (Figure \ref{diploid-well-bridging}).
% The same is shown in Figure \ref{diploid-well-bridging}.

%\hl{Let $c_{\text{LW}}$ denote the amount of coverage required to cover the genome as per the Lander-Waterman coverage analysis} \cite{lander1988genomic}, and let $N_{\text{LW}}$ be the corresponding number of reads. Denote $c$ as the coverage required by an algorithm. The \emph{Normalized coverage depth} ($\bar{c}$) equals the coverage depth $c$ normalized by $c_{\text{LW}}$, i.e., $\bar{c} = \frac{c}{c_{\text{LW}}}$.

Let $c_{\text{LW}}$ be the sequencing coverage depth required to sequence every nucleotide in the genome with a desired probability as per the Lander-Waterman analysis \cite{lander1988genomic}. 
%The corresponding coverage depth, denoted as $c_{\text{LW}}$, is $N_{\text{LW}} L / G$. 
Sequencing the entire genome is necessary, but it does not guarantee that genome reconstruction is possible. 
Let $c$ be the minimum coverage depth required by a genome reconstruction algorithm. 
In our experiments in Section \ref{sec:experiments}, we measure the required \emph{normalized coverage depth} $\bar{c} = \frac{c}{c_{\text{LW}}}$ , i.e., coverage depth $c$ normalized by $c_{\text{LW}}$.

%and let $N_{\text{LW}}$ be the corresponding number of reads. Denote $c$ as the coverage required by an algorithm. The \emph{Normalized coverage depth} ($\bar{c}$) equals the coverage depth $c$ normalized by $c_{\text{LW}}$, i.e., $\bar{c} = \frac{c}{c_{\text{LW}}}$.

%\emph{Normalized coverage depth} ($\bar{c}$) is the ratio of the amount of coverage required by an algorithm (denoted by $c$) for the reconstruction of the genome to $c_{\text{LW}}$. Mathematically, $\bar{c} = \frac{c}{c_{\text{LW}}} = \frac{N}{N_{\text{LW}}}$.}

A diploid genome reconstruction $(\mathcal{G}_0, \mathcal{G}_1)$ is said to be equivalent to the  ground truth genome $(\mathcal{H}_0, \mathcal{H}_1)$ \emph{up to switch errors} if $\exists (\mathcal{G}_0', \mathcal{G}_1')$ such that (a) $\mathcal{G}_0'$ is a rotation of $\mathcal{G}_0$, (b) $\mathcal{G}_1'$ is a rotation of $\mathcal{G}_1$, (c) $\mathcal{G}_0'[i] = \mathcal{G}_1'[i] = \mathcal{H}_0[i]$ for all $i \in \{0, \ldots, \lvert \mathcal{H} \rvert-1\} \setminus \{\mathcal{L}_1, \ldots, \mathcal{L}_n\},$  
%$\mathcal{G}_0'$ and $\mathcal{G}_1'$ differ only at heterozygous loci from $\mathcal{H}_0$ and $\mathcal{H}_1$ respectively
and (d) $\{\mathcal{G}_0'[i], \mathcal{G}_1'[i]\} = \{\mathcal{H}_0[i], \mathcal{H}_1[i]\}$ for all $i \in \{\mathcal{L}_1, \ldots, \mathcal{L}_n\}$. For example, if $(\mathcal{H}_0, \mathcal{H}_1) = (\textsc{A\underline{C}TTT\underline{G}C}, \textsc{A\underline{A}TTT\underline{C}C})$, then a reconstruction $(\mathcal{G}_0, \mathcal{G}_1) = (\textsc{A\underline{A}TTT\underline{G}C}, \textsc{A\underline{C}TTT\underline{C}C})$ is considered equivalent to $(\mathcal{H}_0, \mathcal{H}_1)$ up to switch errors. In this example, the switch error occurs at position $2$. %\hl{These differences in the haplotypes $\mathcal{H}_0, \mathcal{H}_1$ are called single nucleotide variations. Therefore, we seek to achieve two partially phased assemblies $\mathcal{G}_0, \mathcal{G}_1$ that differ only at the positions at which $\mathcal{H}_0, \mathcal{H}_1$ differ.}

%The characters at the heterozygous loci are underlined in this example.
% , while a reconstruction $(\mathcal{G}_0', \mathcal{G}_1') = (\textsc{AATGC}, \textsc{ACTCG})$ is not up to switch errors. 
% The reconstruction is said to be without switch error if it is similar to the ground truth.
    
% \begin{figure*}[hbt!]
%     \centering
%     \includegraphics[scale=0.175]{images/diploid-Well-bridging.png}
%     \caption{\textbf{Figure shows the well-bridging condition for three different arrangements of the maximal double repeat ($r_1, r_2$)}. Only figures towards the right are well-bridged. \textbf{(A1)} and \textbf{(A2)} Both the repeat copies do not cover a heterozygous locus. Figure (A1) has repeat copy $r_1$ bridged by the reads $x_1^1, x_4^1$ and $r_2$ bridged by the read $x_2^2$. Figure (A2) has repeat copy $r_1$ bridged by the reads $x_1^1, x_2^1, x_4^1$ and $r_2$ bridged by the read $x_2^2$. \textbf{(B1)}, \textbf{(B2)}, and \textbf{(B3)} Exactly one repeat copy covers a heterozygous locus. Figure (B1) has repeat copy $r_1$ bridged by the reads $x_1^1, x_4^1$ and $r_2$ unbridged. Figure (B2) has repeat copy $r_1$ bridged by the reads $x_1^1, x_3^1, x_4^1$ and $r_2$ unbridged. Figure (B3) has repeat copy $r_1$ bridged by the reads $x_1^1, x_4^1$ and $r_2$ bridged by the read $x_1^2$. \textbf{(C1)} and \textbf{(C2)} Both repeat copies cover a heterozygous locus. Figure (C1) has repeat copies $r_1$ and $r_2$ unbridged. Figure (C2) has repeat copy $r_1$ bridged by the read $x_1^2$ and $r_2$ unbridged.}
%     \label{diploid-well-bridging}
% \end{figure*}

\subsection{Problem Statement}
Find the minimum read length and coverage depth under which, with high probability (with respect to the sampling distribution of reads), the original haplotype sequences can be reconstructed up to switch errors.\cj{Would a figure with a simple example help the reader here, e.g., given a set of reads, we show (i) true genome, (ii) correct reconstruction up to switch errors, and (iii) an incorrect reconstruction?}

\dm{Complexity of analyzing diploid case over haploid case}

\section{Information theoretic necessary conditions} \label{sec:it}

%  \begin{figure}[hbt!]
%     \centering
%     \includegraphics[scale=0.15]{images/diploid-necessary-condition-proof.png}
%     \caption{Figure shows a set of pair of inter-double repeats $\{\{(r_0^1, r_1^1), (r_0^2, r_1^2)\}, \{(r_0^3, r_1^3), (r_0^4, r_1^4)\}\}$ such that $\lvert s_1 \rvert + \lvert s_2 \rvert = \lvert s_3 \rvert - \lvert s_4 \rvert$, where $s_1, s_2, s_3, s_4$ represent the substrings in the haplotypes. This repeat configuration will be called bridged if at least one of the $8$ copies is bridged.}
%     \label{necessary-condition-proof}
% \end{figure}

These following conditions are necessary to ensure correct reconstruction of the haplotypes up to switch errors.

\label{anchor_uptoerrors}
\begin{enumerate}[label=I\arabic*), ref=I\arabic*]
    \item Intra-triple and intra-interleaved repeats must be bridged.
    % \item \label{information_theoretic_intercondition} A set of pair of inter-double repeats present in the manner shown in Figure \ref{necessary-condition-proof} must be bridged. The case can be generalized to a scenario having $m$ pairs of copies of such repeats where the length of the substrings of the haplotypes present between the pair satisfies the following condition:
    
    % Suppose the repeats are sorted clockwise based on their starting position. Let $s_i (i \in [m])$ denote the substring of the haplotype $\mathcal{H}_0$ from the end of $r_0^{2 \cdot i - 1}$ to the start of $r_0^{2 \cdot i}$ and $s_{i + m} (i \in [m])$ denote the substring of the haplotype $\mathcal{H}_1$ from the end of $r_1^{2 \cdot i - 1}$ to the start of $r_1^{2 \cdot i}$. Now the following must hold :
    % \[\sum_{i = 1}^m c_i \cdot \lvert s_i \rvert = \sum_{i = 1}^m c_{i + m} \cdot \lvert s_{i + m} \rvert\] where $c_i = 1$, if $s_i$ is spelled in the clockwise manner and $c_i = -1$ otherwise \cj{We can consider removing this condition, or simplifying the condition to m=1 case only}.

    \item \label{information_theoretic_intercondition} If there is a pair of inter-double repeats denoted by $\{(s_{0, j_1}^{l_1}, s_{1, j_2}^{l_1}), (s_{0, j_3}^{l_2}, s_{1, j_4}^{l_2})\}$ such that (a) $j_3 - j_1 = j_4 - j_2$, (b) $j_3 > (j_1 + l_1)$, then at least one of the double repeats must be bridged.
    
    \item \label{information_theoretic_covercondition} Reads must cover the genome.
\end{enumerate}

\begin{figure}[hbt!]
    \centering
    \includegraphics[scale=0.16]{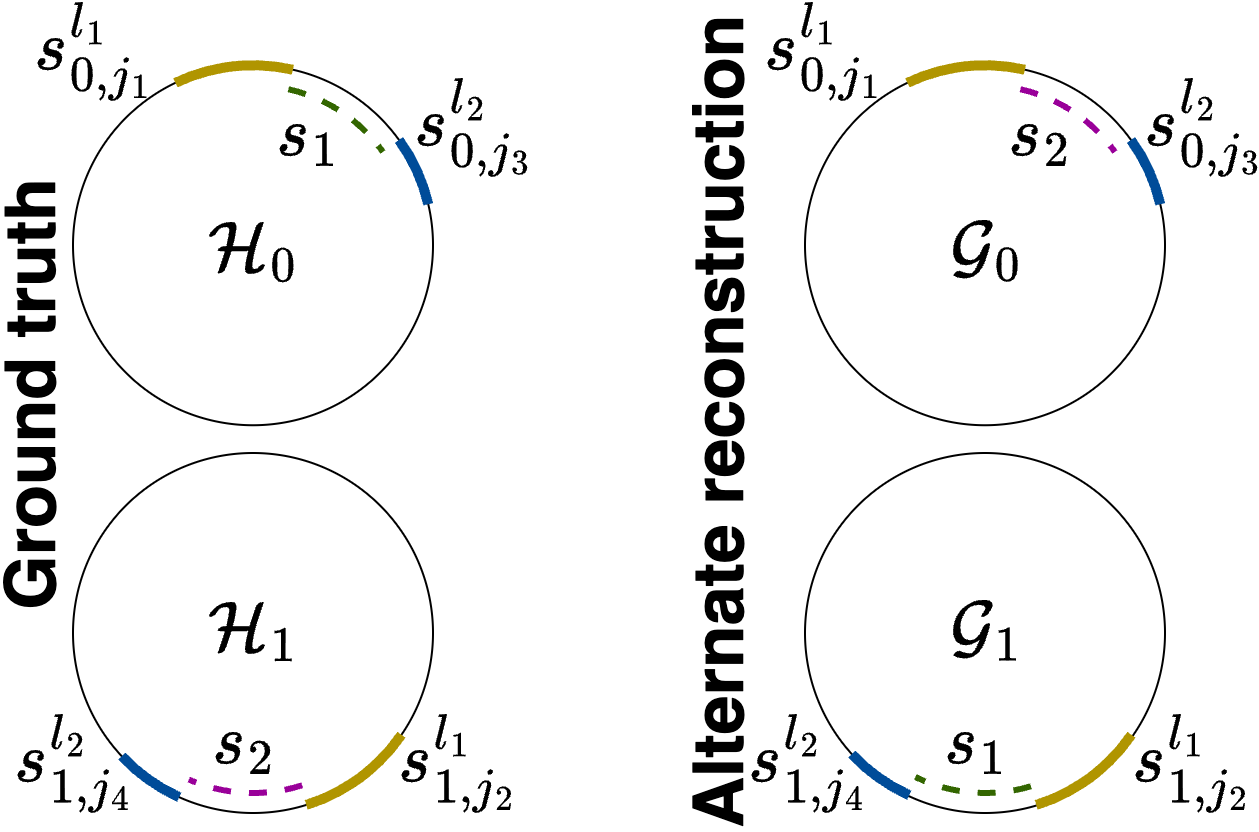}
    \caption{Figure shows a pair of inter-double repeats present in a specific orientation that can lead to an alternate reconstruction if Condition I2 is not satisfied. Here, $s_1$ and $s_2$ represent the substrings between the copies of the yellow and blue repeats in the two haplotypes, respectively. $s_1 \neq s_2$ because of the maximality of the double repeats.}
    \label{diploid-necessary-condition-m=1}
\end{figure}

The violation of Condition I1 during haploid genome assembly leads to two equally likely genome reconstructions, as shown by Shomorony \textit{et al.}~\cite{shomorony2016information}. Accordingly, this condition must be satisfied for diploid genome assembly as well. The necessity of Condition I2 is illustrated using Figure \ref{diploid-necessary-condition-m=1}.  If Condition I2 is not satisfied, that is, neither of the double repeats is bridged, then the likelihood of observing the reads $\mathcal{R}$ is the same for more than one genome; hence, correct reconstruction is not possible.

% \subsection{Without switch errors} \label{anchor_withouterrors}

% \begin{enumerate}
%     \item \ref{anchor_uptoerrors} must hold, i.e.,
    
%     \begin{enumerate}
%         \item Intra-triple and intra-interleaved repeats must be bridged.
%         \item \label{information_theoretic_intercondition} A set of pair of inter-double repeats present in the manner shown in Figure \ref{necessary-condition-proof} must be bridged. The case can be generalized to a scenario having $m$ pairs of copies of such repeats where the length of the substrings of the haplotypes present between the pair satisfies the following condition:
        
%         Suppose the repeats are sorted clockwise based on their starting position. Let $s_i (i \in [m])$ denote the substring of the haplotype $\mathcal{H}_0$ from the end of $r_0^{2 \cdot i - 1}$ to the start of $r_0^{2 \cdot i}$ and $s_{i + m} (i \in [m])$ denote the substring of the haplotype $\mathcal{H}_1$ from the end of $r_1^{2 \cdot i - 1}$ to the start of $r_1^{2 \cdot i}$. Now the following must hold :
%         \[\sum_{i = 1}^m c_i \cdot \lvert s_i \rvert = \sum_{i = 1}^m c_{i + m} \cdot \lvert s_{i + m} \rvert\] where $c_i = 1$, if $s_i$ is spelled in the clockwise manner and $c_i = -1$ otherwise.
    
%         \item Reads must cover the genome.
%     \end{enumerate}
    
%     \item \label{phasingcondition} Suppose the number of heterozygous loci in the genome equals $n$. Then, at least  $n - 1$ adjacent pair of these loci must be covered by a read. This comes from the fact that we need at least $i - 1$ edges to connect $i$ nodes.
% \end{enumerate}

\section{Necessary and sufficient conditions for different algorithms} \label{sec:algo}
\label{algorithmic-necessary-conditions}

In this section, we study the necessary and sufficient conditions for the commonly used assembly algorithms. For each algorithm, the proof of sufficiency involves showing that when all the conditions hold, the algorithm described must return the underlying haplotypes up to switch errors. The proof for the necessity of conditions is done by showing that when any of the conditions fail, some input exists for which the algorithm will fail to produce the desired output.

\subsection{Greedy algorithm}
\subsubsection{Algorithm}
The greedy algorithm (Algorithm \ref{greedy-algorithm}) works by merging pairs of reads in the decreasing order of the length of pairwise overlaps among them. The merged read is added back to the read set, and the process is continued till either the size of the set equals one or no suffix-prefix overlaps are found among reads. Finally, the leftover sequences are processed to obtain two haplotype sequences.
% \subsubsection{Algorithm}
\begin{algorithm}[htpb]
\small
\caption{: Greedy algorithm for diploid genomes}
\label{greedy-algorithm}
\begin{algorithmic}[1]
\State \textbf{Input:} Set of reads $\mathcal{R}$
\State \textbf{Output:} Reconstructed haplotypes $(\mathcal{G}_0, \mathcal{G}_1)$
\State $\mathcal{R}_{tmp} \leftarrow \mathcal{R}$
\While{$\lvert \mathcal{R}_{tmp} \rvert > 1$} 
    \State $x, y \leftarrow \arg \max_{(x, y) : x, y \in \mathcal{R}_{tmp}, x \neq y}\textsc{overlap}(x, y)$

    \If{$\textsc{overlap}(x, y) == 0$}
        \State \textbf{break}
    \EndIf
    
    \State $\mathcal{R}_{tmp} \leftarrow \mathcal{R}_{tmp} \symbol{92} \{x, y\}$

    \State $\mathcal{R}_{tmp} \leftarrow \mathcal{R}_{tmp} \cup \{\textsc{union}(x, y)\}$
\EndWhile

\medskip

% \State \textbf{assert}($\lvert \mathcal{R} \rvert \leq 2$)

% \medskip

\If{$\lvert \mathcal{R}_{tmp} \rvert == 1$}
    \State $\{\mathcal{X}_0\} \leftarrow \mathcal{R}_{tmp}, \  \alpha \leftarrow \textsc{overlap}(\mathcal{X}_0, \mathcal{X}_0)$ \label{line:greedycase1_1}
    
    % \State \textbf{assert}($\textsc{overlap}(\mathcal{G}_1, \mathcal{G}_1) > 0$)

    % \State \textbf{assert}(($\lvert \mathcal{G}_1 \rvert - \textsc{overlap}(\mathcal{G}_1, \mathcal{G}_1)) \% 2 == 0$)
    %\scriptsize
    %\State {\small \textbf{return} $(\mathcal{G}_0, \mathcal{G}_1)$} $\leftarrow (\mathcal{X}_2[: (\lvert \mathcal{X}_2 \rvert - \textsc{overlap}(\mathcal{X}_2, \mathcal{X}_2)) / 2], \mathcal{X}_2[(\lvert \mathcal{X}_2 \rvert - \textsc{overlap}(\mathcal{X}_2, \mathcal{X}_2)) / 2 : \lvert \mathcal{X}_2 \rvert - \textsc{overlap}(\mathcal{X}_2, \mathcal{X}_2)])$ 
    \small
    \State \textbf{return} $(\mathcal{G}_0, \mathcal{G}_1) \leftarrow (\mathcal{X}_0[: (\lvert \mathcal{X}_0 \rvert - \alpha) / 2], \mathcal{X}_0[(\lvert \mathcal{X}_0 \rvert -\alpha) / 2 : \lvert \mathcal{X}_0 \rvert - \alpha])$ \label{line:greedycase1_2}
\Else
    \State $\{\mathcal{Y}_0, \mathcal{Y}_1\} \leftarrow \mathcal{R}_{tmp}$
    
    % \State \textbf{assert}($\textsc{overlap}(\mathcal{G}_1, \mathcal{G}_1) > 0$ \textbf{and} $\textsc{overlap}(\mathcal{G}_2, \mathcal{G}_2) > 0$)

    % \State \textbf{assert}($\lvert \mathcal{G}_1 \rvert - \textsc{overlap}(\mathcal{G}_1, \mathcal{G}_1) == \lvert \mathcal{G}_2 \rvert - \textsc{overlap}(\mathcal{G}_2, \mathcal{G}_2)$)
    
    \State \textbf{return} $(\mathcal{G}_0, \mathcal{G}_1) \leftarrow (\mathcal{Y}_0[: \lvert \mathcal{Y}_0 \rvert - \textsc{overlap}(\mathcal{Y}_0, \mathcal{Y}_0)], \mathcal{Y}_1[: \lvert \mathcal{Y}_1 \rvert - \textsc{overlap}(\mathcal{Y}_1, \mathcal{Y}_1)])$ 
\EndIf

\end{algorithmic}
\end{algorithm}

\subsubsection{Necessary and sufficient conditions for correct reconstruction}
\label{greedy_withoutphasing}

\begin{enumerate}[label=G\arabic*), ref=G\arabic*]

    \item \label{condition0_greedy} The information-theoretic necessary conditions stated in Section \ref{anchor_uptoerrors} must hold.
    \item \label{condition1_greedy} The number of heterozygous loci in the genome must be at least $2$.
    \item \label{condition2_greedy} Both the following conditions must hold:
    
    \begin{enumerate}
        \item \label{condition2.1_greedy} Read $x_1^i$ must overlap with reads $x_3^i$ and  $x_4^i$, $\forall i \in [n]$. Read  $x_2^i$ must overlap with reads $x_3^i$ and $x_4^i$, $\forall i \in [n]$. Note that the reads $x_1^i$ (resp. $x_2^i$) and $x_3^i$ (resp. $x_4^i$) may be equal, in which case an overlap between $x_1^i$ (resp. $x_2^i$) and $x_3^i$ (resp. $x_4^i$) is not necessary.
        
        \item \label{condition2.2_greedy} Double repeats must be well-bridged.
    \end{enumerate}
\end{enumerate}

% The necessity of these conditions is proved in Appendix Section \ref{greedynecessity}.

\subsubsection{Proof of the necessity of Conditions~\ref{condition1_greedy} and \ref{condition2_greedy}}
\label{greedynecessity}
\begin{Claim}
    Condition \ref{condition1_greedy} is necessary for unique reconstruction by Algorithm \ref{greedy-algorithm}. 
\end{Claim}

\begin{proof}
    \begin{figure}[hbt!]
        \centering
        \includegraphics[scale=0.12]{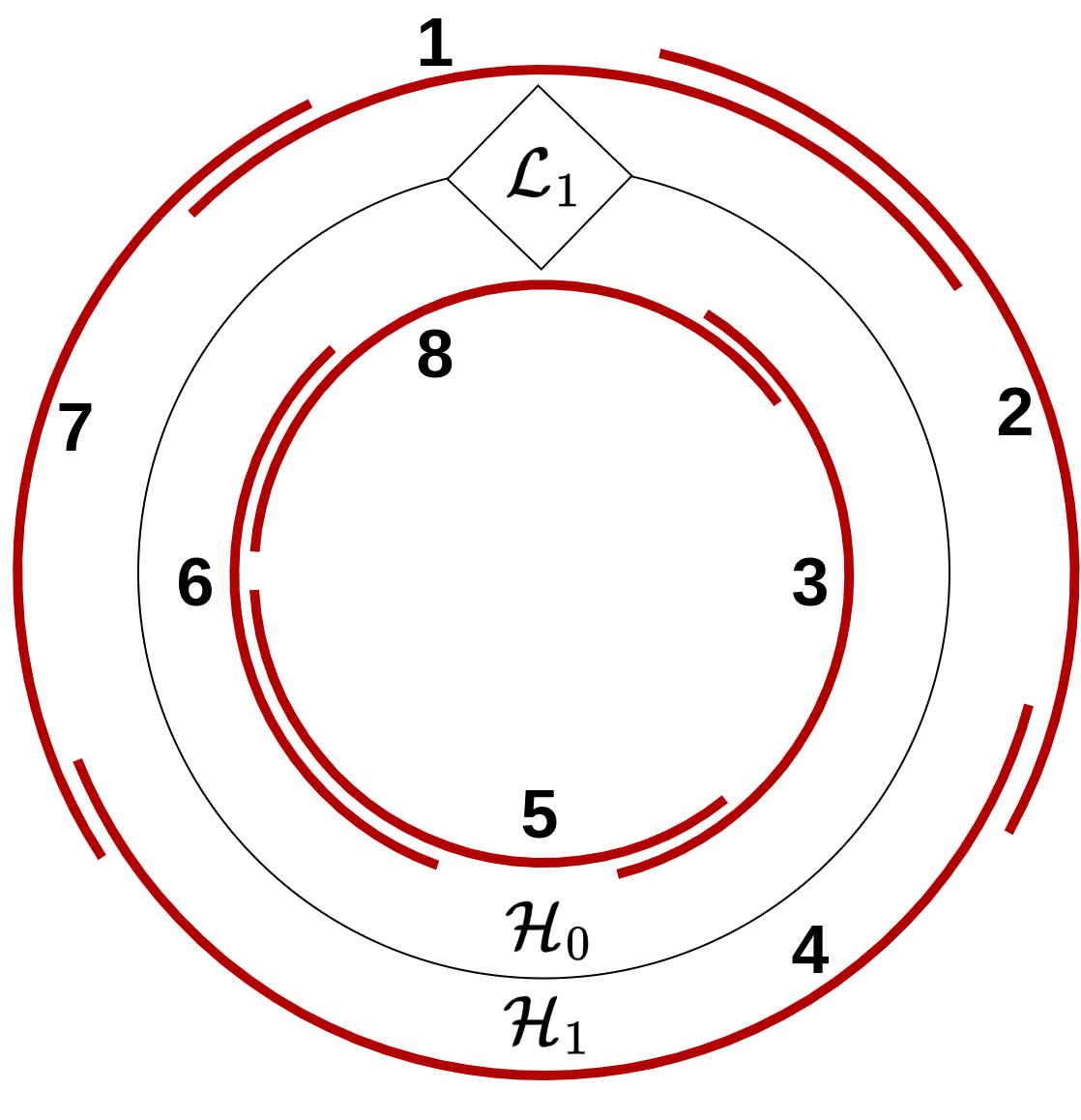}
        \caption{Reads sampled from a diploid genome containing a single heterozygous locus.}
        \label{diploid-greedy-n=1}
    \end{figure}

    Suppose $\mathcal{G}$ has a single heterozygous locus, and the reads are sampled as shown in Figure \ref{diploid-greedy-n=1}. At the end of the while loop, the greedy algorithm (Algorithm \ref{greedy-algorithm}) produces a chain that links the reads in the order numbered $1$ to $8$. Subsequently, Algorithm \ref{greedy-algorithm} (Lines \ref{line:greedycase1_1}-\ref{line:greedycase1_2}) does not return the correct genome.
    %because read $1$ and read $8$ cover the heterozygous locus on opposite haplotypes and do not overlap.
    %check for approximate suffix-prefix overlaps between read $1$ and read $8$.
    % The only way to get a unique solution is to have reads of length strictly greater than $\lvert \mathcal{H} \rvert$, which is not possible.\cj{consider removing the last sentence} 
\end{proof}

\emph{Remark:} In the above example, one could consider modifying the greedy algorithm such that it looks for an approximate match between read $8$ and read $1$ varying in exactly one position. However, such overlaps may not be unique. For example, if a prefix of read $1$ is $s_1 + x + s_2 + z + s_1 + y + s_2$ and a suffix of read $8$ is $s_1 + x + s_2 + w + s_1 + y + s_2$ (where $s_1, s_2$ are strings of positive length and $x, y, z, w$ are characters belonging to $\{\textsc{A}, \textsc{C}, \textsc{T}, \textsc{G}\}$), there are at least two possible overlaps of lengths $\lvert s_1 \rvert + \lvert s_2 \rvert + 1$ and $2 \cdot (\lvert s_1 \rvert + \lvert s_2 \rvert) + 3$ which have a hamming distance of one. 

\begin{Claim} \label{claim2_greedy}
    Condition \ref{condition2_greedy} is necessary for unique reconstruction by Algorithm \ref{greedy-algorithm}. 
\end{Claim}

\begin{proof}
    
    \begin{figure}[hbt!]
        \centering
        \includegraphics[scale=0.2]{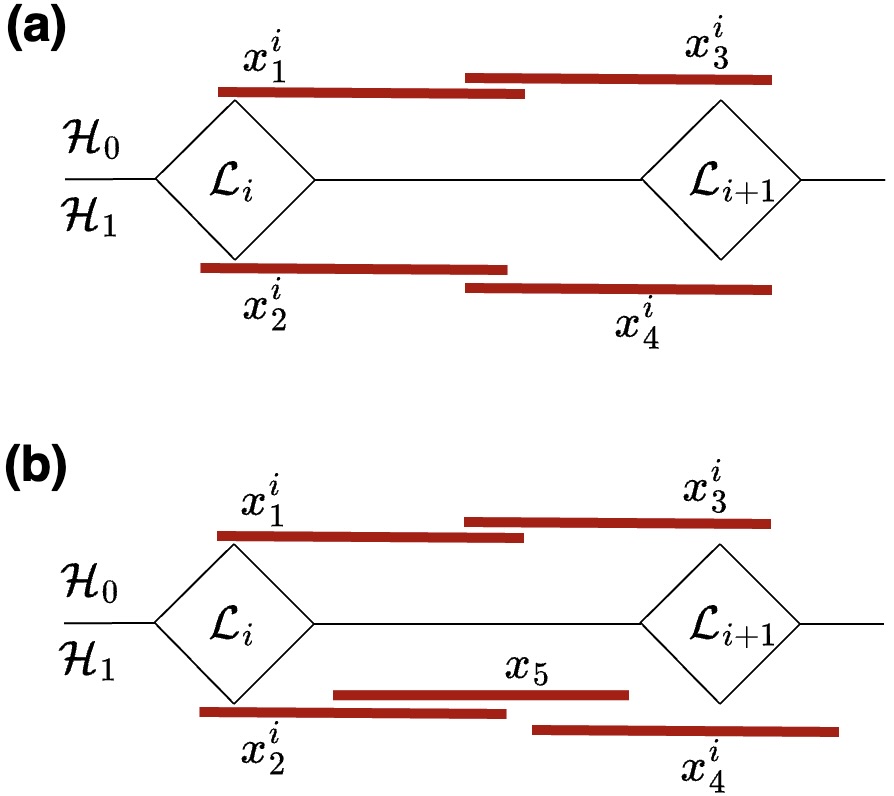}
        \caption{\textbf{(a)} The set of reads satisfying condition \ref{condition2.1_greedy}. \textbf{(b)} The set of reads not satisfying condition \ref{condition2.1_greedy} because read $x_2^i$ does not overlap with read $x_4^i$. }
        \label{diploid-greedy-3a}
    \end{figure}
        
    If Condition \ref{condition2.1_greedy} is not satisfied, then as depicted in Figure \ref{diploid-greedy-3a}(b), read $x_1^i$ can first get merged with read $x_5$ which further merges with read $x_3^i$, thereby leaving no chance for $x_2^i$ to merge with $x_4^i$ and hence resulting in a gap in the reconstruction. Next, we consider Condition \ref{condition2.2_greedy} which requires all double repeats to be well-bridged. We prove that this condition is necessary irrespective of the arrangement of a double repeat. A double repeat can occur in various forms, for example, the coordinates of repeat copies may or may not overlap with each other. Similarly, either zero or one or both repeat copies may cover a heterozygous locus. In total, we consider all nine  arrangements.
    % For all cases involving double repeats, unique reconstruction up to switch errors also demands these repeats to be well-bridged. 
    Figure \ref{greedy-extra-condition} in Appendix shows that for all these cases, correct genome reconstruction is not possible if the double repeat is not well-bridged.
    % \DM{This involves doing an exhaustive case analysis  and includes the following $9$ cases: $(1)$ and $(2)$ Both the repeat copies do not cover a heterozygous locus and lie within the adjacent pair of heterozygous loci, and the two copies overlap (or do not overlap); $(3)$ Both the repeat copies do not cover a heterozygous locus and do not lie within the adjacent pair of heterozygous loci; $(4)$ Exactly one copy covers the heterozygous locus, and the two copies overlap; $(5)$ and $(6)$ Exactly one copy covers the heterozygous locus (suppose first) and the two copies do not overlap and the second copy lie (or do not lie) in the homozygous region that is adjacent to the heterozygous loci covered by the first copy; $(7)$ Both copies cover the heterozygous locus but do not cover any common heterozygous locus; $(8)$ and $(9)$ Both copies have at least one loci in common and belong to same (or different) haplotypes.} 
\end{proof}

\subsubsection{Proof for the sufficiency of Conditions~\ref{condition0_greedy}--\ref{condition2_greedy}}

% \begin{Claim}
% \label{greedymovement}
% After removing the restriction $x \neq y$ on line $2$ in algorithm \ref{greedy-algorithm}, condition \ref{condition2_greedy} ensures that $\forall R_1 \in \mathcal{R}$ such that $R_1$ covers a loci $\mathcal{L}_i$, $\exists R_2 \in \mathcal{R}$ such that $R_2$ covers loci $\mathcal{L}_{i + 1}$ and the following two conditions hold:

% \begin{enumerate}
%     \item $R_1$ and $R_2$ belong to the same string $s$ output by the greedy algorithm.
%     \item String $s$ formed by the concatenation of the reads $R_1$ to $R_2$ in the order they appear in $s$ covers the region between loci $\mathcal{L}_{i}$, $\mathcal{L}_{i + 1}$.
% \end{enumerate}
% \end{Claim}

% \begin{proof}
% Condition \ref{condition2.1_greedy} ensures that the claim is valid if $s$ doesn't cover any repeat region. In case, it does cover, well-bridging of the double repeats will ensure that greedy behaves the same way if there were no repeats. 
% \end{proof}
\begin{figure}[htpb]
    \centering
    \includegraphics[scale=0.15]{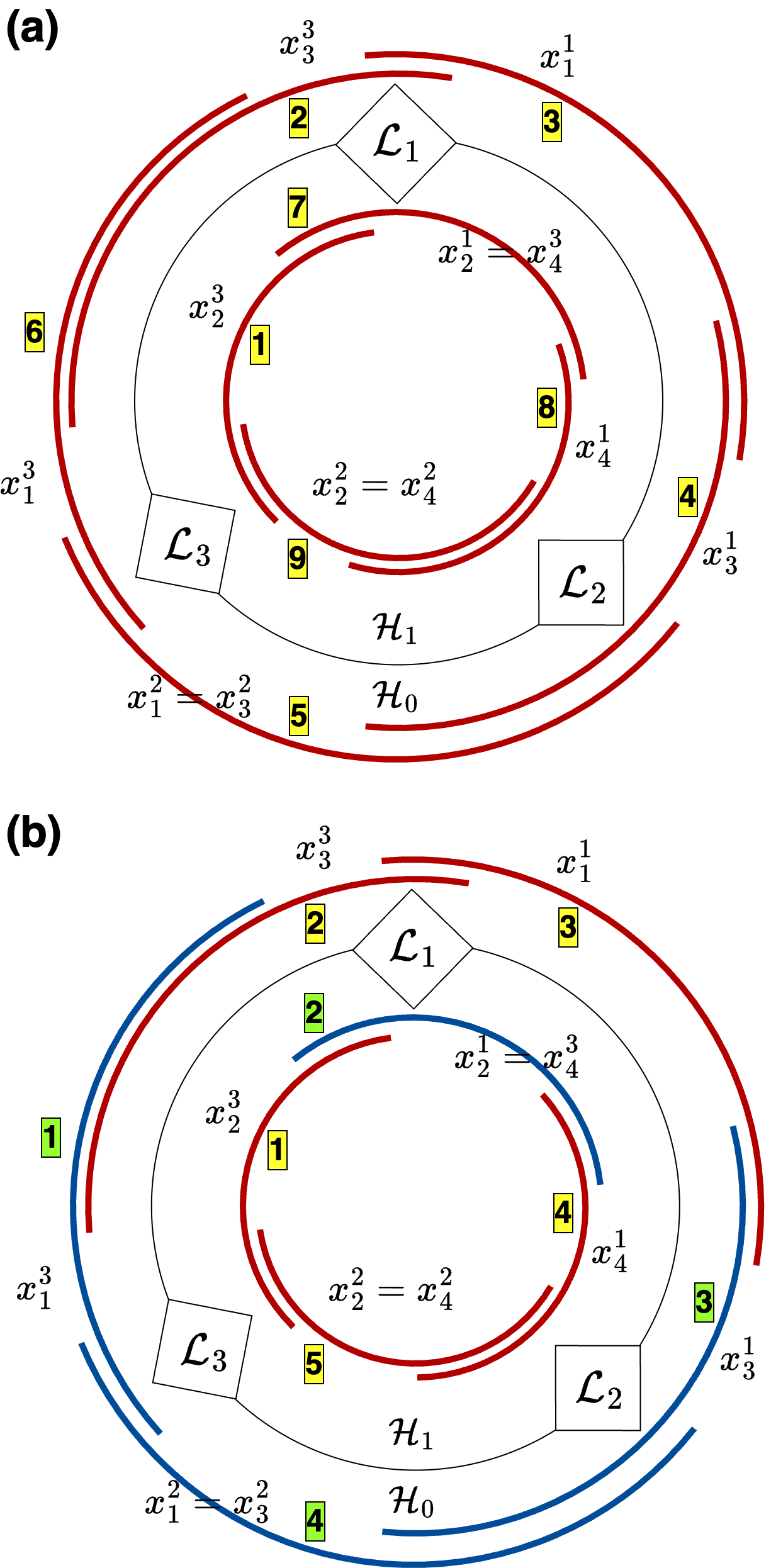}
    \caption{Figure shows the output of the greedy algorithm under two different sets of reads as input. \textbf{(a)} The output is a single string composed of reads $1$ to $9$ in order. \textbf{(b)} The output comprises two strings, one formed from the red-colored reads and the other from the blue-colored reads.}
    \label{diploid-greedy-sufficiency}
\end{figure}

\begin{Claim}
The number of strings left at the end of the while loop in the greedy algorithm cannot be greater than $2$.%\cj{do you want to refer to the `end', or would it be better to make this claim after the algorithm completes the while loop.}
\end{Claim}

\begin{proof}

Let $\mathcal{R'}$ be the maximal subset of $\mathcal{R}$ such that the reads belonging to $\mathcal{R}'$ do not cover a heterozygous locus. Condition \ref{condition2_greedy} ensures that when the greedy algorithm is run with $\mathcal{R} \symbol{92} \mathcal{R}'$ as input, we will get, at most, two directed open chains of reads at the end of the while loop. 
%These chains are sufficient to recover the true haplotypes up to switch errors. 
The output size will be one if an odd number of switch errors are made in the underlying haplotypes; otherwise, the output size will be two  (refer to Figure \ref{diploid-greedy-sufficiency} for an example). What is left to be proved is that the reads belonging to $\mathcal{R}'$ would only result in the subdivision\footnote{A directed edge subdivision is the insertion of a new vertex $v_j$ in the middle of an existing directed edge $e = (v_i, v_k)$, hence replacing it with two new directed edges $e^{'} = (v_i, v_j)$ and $e^{''} = (v_j, v_k)$. A directed graph subdivision is a sequence of directed edge subdivisions.} and extension of the output chains for the input $\mathcal{R} \symbol{92} \mathcal{R}'$ and hence will not result in the formation of any additional chains. 

Suppose the reads in $\mathcal{R}$ and $\mathcal{R}'$ are sorted in clockwise order based on their starting positions (hence based on their ending positions as well). Consider the minimum index $a$, and then, the maximum index $b$ ($a, b \in [\lvert \mathcal{R}' \rvert]$) such that $\textsc{overlap}(\mathcal{R}'[i], \mathcal{R}'[i + 1]) > 0, \forall i \in [a, b)$. Next, consider the read $u$ just before read $\mathcal{R}'[a]$ in $\mathcal{R}$ and the read $v$ just after read $\mathcal{R}'[b]$ in $\mathcal{R}$. The only way the greedy algorithm skips reads $\mathcal{R}'[a], \ldots, \mathcal{R}'[b]$ is if it merges read $u$ with read $v$. But since $\textsc{overlap}(u, \mathcal{R}'[a]) > \textsc{overlap}(u, v)$ and $\textsc{overlap}(\mathcal{R}'[b], v) > \textsc{overlap}(u, v)$, the greedy algorithm will merge the reads in the order $u \rightarrow \mathcal{R}'[a] \rightarrow \ldots \rightarrow \mathcal{R}'[b] \rightarrow v$ (note that $\textsc{overlap}(u, \mathcal{R}'[a]) > 0$ by virtue of Condition \ref{information_theoretic_covercondition}). We can extend this argument for indices greater than $b$ and keep doing the same till $\mathcal{R}'$ is exhausted. This proves the claim.
\end{proof}

The above claim confirms that the output at the end of the while loop consists of, at most, two strings. Next, we analyze the case when the output is a single string. Suppose $i_1, i_2, \ldots, i_{\lvert \mathcal{R} \rvert}$ are the IDs of the reads (w.r.t. set $\mathcal{R}$) 
in the order in which they appear in the output string $\mathcal{X}_0$.
%(considered from left to right). 

\begin{Claim} \label{claim_endreadscover}
Reads $i_1$ and $i_{\lvert \mathcal{R} \rvert}$ must each cover a heterozygous locus in the underlying haplotypes.
%, and similarly, read $i_{\lvert \mathcal{R} \rvert}$ must cover a heterozygous locus.    
\end{Claim}

\begin{figure}[htbp]
    \centering
    \includegraphics[scale=0.34]{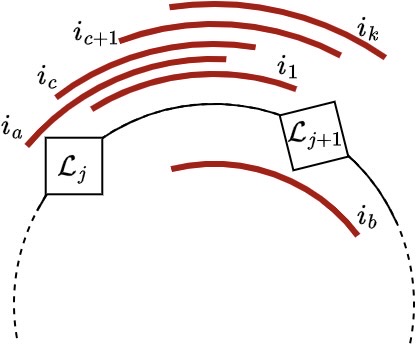}
    \caption{Figure illustrating the argument in Claim \ref{claim_endreadscover}.}
    \label{claim2-image}
\end{figure}

\begin{proof}
 Suppose read $i_1$ does not cover a heterozygous locus. Let $k$ be the minimum index such that read $i_k$ covers a heterozygous locus $\mathcal{L}_{j + 1}$. Note that $i_1$ must belong to the region between the loci pair $(\mathcal{L}_{j}, \mathcal{L}_{j + 1})$. Let us analyze the reads $x_1^j, x_2^j, x_3^j, x_4^j$. By definition, it is clear that $i_k$ is either $x_3^j$ or $x_4^j$. WLOG, let it be $x_3^j$. Now, since the output is a single string formed by the iterative merging of all the reads (including $x_3^j$ and $x_4^j$), there exists a contiguous sequence of reads $i_a, i_{a + 1}, \ldots, i_b,$ in the output string where $k < a \leq b < \lvert \mathcal{R} \rvert$ and $i_a = x_1^j \text{ or } x_2^j$ and $i_b = x_4^j$. Since the endpoint of $i_1$ is after the endpoint of $i_a$ and the start point of $i_1$ is before the start point of $i_b$, as a result of Condition \ref{information_theoretic_covercondition} there exists some index $c \in [a, b)$, such that $\textsc{overlap}(i_c, i_1) > \textsc{overlap}(i_c, i_{c + 1})$, which is a contradiction (since $i_1$ will not then be the first read in the output). See Figure \ref{claim2-image} for an illustration. A similar argument can be used to prove the claim for read $i_{\lvert \mathcal{R} \rvert}$. 
\end{proof}

% Applying this result with $R_1 = i_{\lvert \mathcal{R} \rvert}$ proves the claim.

\begin{Claim} \label{gapclaim}
% Assume $i_{\lvert \mathcal{R} \rvert}$ and $i_1$ do not cover a common heterozygous locus. 
Let $\mathcal{L}_p$ (resp.\ $\mathcal{L}_q$) be the heterozygous locus covered by read $i_{\lvert \mathcal{R} \rvert}$ (resp.\ read $i_1$) that is furthest, in the clockwise (resp.\ anti-clockwise) direction, from the starting (resp.\ ending) position of the read. 
%Let $\mathcal{L}_q$ be the heterozygous locus covered by read $i_1$ that is furthest, in the anti-clockwise direction from the ending position of the read. 
If $i_{\lvert \mathcal{R} \rvert}$ and $i_1$ do not cover a common heterozygous locus, then $i_{\lvert \mathcal{R} \rvert} \in \{x_1^p, x_2^p\}$ and $i_1 \in \{x_3^{q - 1}, x_4^{q - 1}\}$.
\end{Claim}

% \begin{proof}
% \begin{figure}[hbt!]
%     \centering
%     \includegraphics[scale=0.2]{images/diploid-proof-image.png}
%     \caption{Figure shows the case when $\textsc{overlap}(r', r) = \textsc{overlap}(i_{\lvert \mathcal{R} \rvert}, r)$ and the bold yellow region represent a double repeat.}
%     \label{diploid-proof-image}
% \end{figure}

\begin{figure}[htbp]
    \centering
    \includegraphics[scale=0.21]{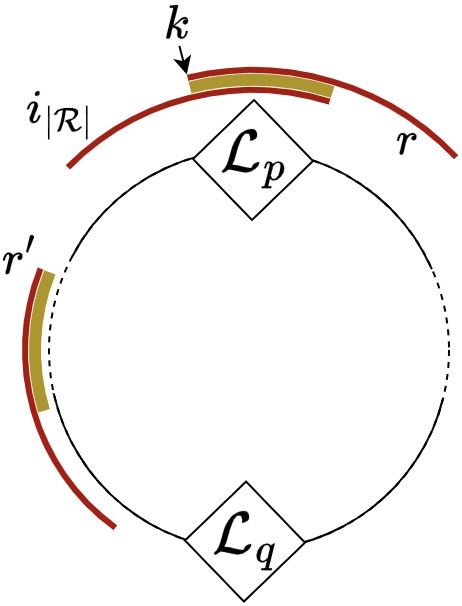}
    \caption{Figure shows the case when $\textsc{overlap}(r', r) = \textsc{overlap}(i_{\lvert \mathcal{R} \rvert}, r)$ and the bold yellow substring represents a double repeat.}
    \label{diploid-proof-image}
\end{figure}

\begin{proof}
% \hl{If the claim does not hold, there exists a double repeat of length at least $L - 1$, which is not possible under condition} \ref{condition2.2_greedy}. 
We prove the result by contradiction for read $i_{\lvert \mathcal{R} \rvert}$, and the same can also be extended for read $i_1$. WLOG, let us assume that $i_{\lvert \mathcal{R} \rvert}$ belongs to haplotype $\mathcal{H}_0$. Suppose for the sake of contradiction that $i_{\lvert \mathcal{R} \rvert} \neq  x_1^p$. Consider all the reads that cover $\mathcal{L}_p$, and choose the one which has the longest prefix matching with a suffix of $i_{\lvert \mathcal{R} \rvert}$. 
%which has maximum prefix-suffix overlap with $i_{\lvert \mathcal{R} \rvert}$. 
Let it be $r = s_{0,k}^L$, where $k \in [0, \lvert \mathcal{H} \rvert - 1]$. 

Now, since $i_{\lvert \mathcal{R} \rvert}$ is the last read in the string output by the greedy algorithm, therefore, $\exists r' \in \mathcal{R} \symbol{92} \{r\}$ such that $\textsc{overlap}(r', r) \geq \textsc{overlap}(i_{\lvert \mathcal{R} \rvert}, r)$ and greedy merges $r'$ with $r$ (else, the greedy algorithm would have merged $i_{\lvert \mathcal{R} \rvert}$ with $r$, contradicting the fact that $i_{\lvert \mathcal{R} \rvert}$ is the last read in $\mathcal{X}_0$). Also $r' \neq s_{0,k + \textsc{overlap}(r',r) - L}^L$, else $r'$ will have a greater prefix-suffix overlap with $i_{\lvert \mathcal{R} \rvert}$ as compared to $r$ (see Figure \ref{diploid-proof-image}). This implies there exists a double repeat $s_{0, k}^{\textsc{overlap}(r', r)}$ (one in $\textsc{union}(i_{\lvert \mathcal{R} \rvert}, r)$, and other in $r'$). The double repeat cannot be bridged in $\textsc{union}(i_{\lvert \mathcal{R} \rvert}, r)$ because of the maximality of the overlap of $r$ with $i_{\lvert \mathcal{R} \rvert}$. Suppose it is bridged in $r'$, by some read $r''$ such that $r''$ has maximum prefix-suffix overlap with $r'$. This implies $\textsc{overlap}(r', r'') > \textsc{overlap}(r', r)$, but since the greedy algorithm merges $r'$ with $r$, this implies $\exists r''' \in \mathcal{R} \symbol{92} \{r'\}$ such that $\textsc{overlap}(r''', r'') \geq \textsc{overlap}(r', r'')$ and the greedy algorithm merges $r'''$ with $r''$. This implies another double repeat of length $\textsc{overlap}(r''', r'')$ (one in $\textsc{union}(r', r'')$, and the other in $r'''$) which is not bridged in $\textsc{union}(r', r'')$ and therefore has to be bridged in $r'''$. Note that $\textsc{overlap}(r''', r'') > \textsc{overlap}(r', r)$. The argument can be extended till the length to be bridged exceeds $L - 2$, which cannot be bridged by any read of length $L$. Hence, we reach a contradiction. 
\end{proof}

\begin{Claim} \label{claim_endreadsoverlap}
$i_{\lvert \mathcal{R} \rvert}$ has a suffix-prefix overlap with $i_1$.
\end{Claim}

\begin{proof}
We consider two cases.
%\begin{itemize}
    \emph{Case (1) $i_{\lvert \mathcal{R} \rvert}$ and $i_1$ do not cover a common heterozygous locus.} From Claim \ref{gapclaim}, we know that $i_{\lvert \mathcal{R} \rvert} \in \{x_1^p, x_2^p\}$ and $i_1 \in \{x_3^{q - 1}, x_4^{q - 1}\}$. We further argue that $p = q - 1$, else the region from the end of $i_{\lvert \mathcal{R} \rvert}$ to the start of $i_1$ is ignored in the reconstruction of one of the haplotypes. Condition \ref{condition2.1_greedy} ensures that $i_{\lvert \mathcal{R} \rvert}$ and $i_1$ overlap.
    % Therefore, $i_{\lvert \mathcal{R} \rvert} \in \{x_1^p, x_2^p\}$ and $i_1 \in \{x_3^{p}, x_4^{p}\}$.
    \emph{Case (2) $i_{\lvert \mathcal{R} \rvert}$ and $i_1$ cover a common heterozygous locus.} $i_{\lvert \mathcal{R} \rvert}$ and $i_1$ must cover the same haplotype, else again the region from the end of $i_{\lvert \mathcal{R} \rvert}$ to the start of $i_1$ is ignored in a haplotype.
    % If not, then Claim \ref{gapclaim} can be extended for this case, and as a result, the while loop cannot terminate without merging $i_{\lvert \mathcal{R} \rvert}$ with either $x_3^p$ or $x_4^p$. Hence $i_{\lvert \mathcal{R} \rvert}$ will not be the last read in the output string, which contradicts our assumption.
%\end{itemize}
\end{proof}

\begin{Claim} \label{claim_removeoverlap}
    % Suppose the greedy algorithm outputs a single string $\mathcal{G}_0$ at the end of the while loop. Then, r
    Removing the longest proper suffix of $\mathcal{X}_0$ that matches a prefix of $\mathcal{X}_0$ gives a string that is a concatenation of the two haplotypes up to switch errors.
\end{Claim}

\begin{proof}
    Observe that the maximum length of an overlap of $\mathcal{X}_0$ to itself is strictly less than $L$. If not, then in this overlap, the start index of read $i_{\lvert \mathcal{R} \rvert}$ lies ahead of the end index of read $i_1$ when considered clockwise. This implies $\exists c \in [1, \lvert \mathcal{R} \rvert - 1)$ such that $\textsc{overlap}(i_{\lvert \mathcal{R} \rvert}, i_{c + 1}) > \textsc{overlap}(i_c, i_{c + 1})$, which is not possible, since then $i_{\lvert \mathcal{R} \rvert}$ will not be the last read in the output string.

    Suppose there are two possible proper overlaps of $\mathcal{X}_0$ to itself. Let us denote the strings corresponding to the two overlaps as $s_1, s_2$, where $0 < \lvert s_1 \rvert < (\lvert s_2 \rvert = \textsc{overlap}(i_{\lvert \mathcal{R} \rvert}, i_1)) < L$.  
    %Therefore $s_2[: \lvert s_1 \rvert] = s_2[\lvert s_2 \rvert - \lvert s_1 \rvert :] = s_1$. 
    % From Claim \ref{gapclaim} and Claim \ref{claim_endreadsoverlap}, if $s_1$ was the correct overlap, that would leave string $s_2$ as a double repeat that is not well-bridged, which is a contradiction (\hl{That $s_2$ is not well-bridged can be seen by examining two cases: (1) $s_2$ covers a heterozygous locus, and (2) $s_2$ does not cover a heterozygous locus. In both cases, if $s_2$ was well-bridged, that would imply that either $i_1$ is not the first read in the output string or $i_{\lvert \mathcal{R} \rvert}$ is not the last read in the output string}). 
    If $s_2$ is not the correct overlap, that would imply the existence of a double repeat (one copy covered by $i_1$ and the other covered by $i_{\lvert \mathcal{R} \rvert}$) which is not well-bridged. This statement can be proved via contradiction as following. We have the following two cases: 
    
    \noindent (1) Suppose there is a copy of the repeat that covers a heterozygous locus and is bridged. WLOG assume such a case exists w.r.t. the repeat copy belonging to read $i_1$. Consider the read $r$ that bridges this repeat and has maximum overlap with $i_1$. Clearly $r \neq i_{\lvert \mathcal{R} \rvert}$, else $\textsc{overlap}(i_{\lvert \mathcal{R} \rvert}, i_1) > \lvert s_2 \rvert$. Also, $\textsc{overlap}(r, i_1) > \max_{r' \in \mathcal{R} \symbol{92} r}\textsc{overlap}(r, r')$, else $r$ will not both bridge and have maximum overlap amongst all the reads that do bridge the repeat. This implies $r$ will be merged with $i_1$ in some step of the greedy algorithm, and hence $i_1$ will not be the terminal read as assumed before. 
    
    \noindent (2) Suppose at least one copy of the repeat does not cover a heterozygous locus and is double-bridged by reads sampled from opposite haplotypes covering a common heterozygous locus. The first half of the statement implies that the terminal reads do not share a common heterozygous locus and the second half implies that at least one of the pair in the set $\{\{x_1^p, x_2^p\}, \{x_3^{q - 1}, x_4^{q - 1}\}\}$ must well-bridge the repeat. Next, it is easy to see from Claim \ref{gapclaim} that these are self-contradicting statements. This proves that removing the maximum overlap is a correct operation. 
    % Next, we show why the output string is sufficient to extract the concatenation of the two haplotypes up to switch errors. 

    % Denote ${\mathcal{L}_k}^i$ as the character at $k^{th}$ heterozygous locus in haplotype $i$. Suppose the order in which string $\mathcal{G}_0$ covers these characters is ${\mathcal{L}_k}^{i_1}, {\mathcal{L}_{k + 1}}^{i_2}, \ldots, {\mathcal{L}_n}^{i_{1 + n - k}}, {\mathcal{L}_1}^{i_{2 + n - k}}, \ldots, {\mathcal{L}_{k - 1}}^{i_n}, {\mathcal{L}_k}^{1 - i_1},$ 
    
    % ${\mathcal{L}_{k + 1}}^{1 - i_2}, \ldots, {\mathcal{L}_n}^{1 - i_{1 + n - k}}, {\mathcal{L}_1}^{1 - i_{2 + n - k}}, \ldots, {\mathcal{L}_{k - 1}}^{1 - i_n}, {\mathcal{L}_k}^{i_1}$. 
    After removing the correct overlap, the output string can be visualized as the traversal of the heterozygous loci twice, covering a different base of each heterozygous locus in both traversals. The length of the string is $2 \lvert \mathcal{H} \rvert$. This completes the proof.
\end{proof}

The above claims prove the correctness of the greedy algorithm when it outputs a single string. Next, assume that we get two strings as output. This condition arises when there is an even number of switch errors during reconstruction. Claim \ref{claim_endreadsoverlap} can be extended for this case. Hence, each of the two strings has a suffix-prefix overlap with itself. Removal of the maximum proper overlap between the suffix and prefix of $\mathcal{Y}_0$ and $\mathcal{Y}_1$ being a valid operation is a result of Claim \ref{claim_removeoverlap}. After removing the overlapping part, necessary conditions ensure that the two strings are of equal length.

\subsection{De-Bruijn graph}
\subsubsection{Algorithm}
\label{condense-algorithm}

First, an edge-centric de Bruijn graph \cite{chikhi2013space} is built using reads $\mathcal{R}$ and is denoted by $\textsc{DB}_{ec}^k(\mathcal{R})$. This is a graph in which the nodes are the set of all $k$-length substrings ($k$-mers) occurring in $\mathcal{R}$, $k<L$, and there is an edge from node $u$ to node $v$ iff $\textsc{overlap}(u, v) = k - 1$ and $\textsc{union}(u, v)$ is a substring of some read in $\mathcal{R}$. Next, all maximal unitigs\footnote{Maximal paths in $\textsc{DB}_{ec}^k(\mathcal{R})$ with all but the first vertex having in-degree $1$ and all but the last vertex having out-degree $1$.} are compacted into a single vertex. Since the sequence represented by some nodes is now longer than $k$, this is no longer a $k$-mer graph and is denoted by the \emph{condensed sequence graph} $G_c$. 

If all pairs of adjacent heterozygous loci are separated by at most $k - 1$ nucleotides and $k$ is greater than the maximum length of an inter-double repeat, then $G_c$ will contain two connected components representing individual haplotypes. In this case, the subsequent conditions for the successful reconstruction of the two haplotypes will be analogous to the ones for haploid genomes \cite{bresler2013optimal}. Therefore, our further analysis assumes that $G_c$ is composed of a single connected component. We aim to find the Eulerian circuit of $G_c$ such that the string spelled by the circuit represents the concatenation of two haplotypes up to switch errors. The string spelled by the walk is computed by beginning from the label of a starting vertex and appending the label of the other vertices (excluding the ending vertex in the walk, which is the same as the starting vertex) in the order they are visited after removing the $k - 1$ length prefix of those labels. Finally, the $k - 1$ length suffix of the string obtained is trimmed since that is covered in the label of the starting vertex (Algorithm \ref{dbg-algorithm}). 

\begin{algorithm}[htpb]
\small
\caption{: Getting haplotypes up to switch errors from $G_c$}
\label{dbg-algorithm}
\begin{algorithmic}[1]

\State \textbf{Input:} Condensed sequence graph $G_c$
\State \textbf{Output:} Reconstructed haplotypes $(\mathcal{G}_0, \mathcal{G}_1)$
\State $A :$ Array where $A[i]$ stores the label of the vertex indexed $i$ in $G_c$.
\State $\textsc{ETOUR} :$ List of vertices in the order visited by an Eulerian tour on $G_c$, where all vertices are distinct (the starting vertex is added only once).

%\medskip

\State $S \leftarrow A[\textsc{ETOUR}[0]]$

%\medskip
\For{$i = 1 \text{ to } \lvert \textsc{ETOUR} \rvert - 1$}
    \State $S \leftarrow S + A[\textsc{ETOUR}[i]][k - 1 :]$
\EndFor

%\medskip

\State $S \leftarrow S[: \lvert S \rvert - k + 1]$

% \State \textbf{assert}($\lvert S \rvert \% 2 == 0$)

%\medskip

\State return $(S[: \lvert S \rvert / 2], S[\lvert S \rvert / 2 :])$

\end{algorithmic}
\end{algorithm}

\subsubsection{Sufficient conditions for correct reconstruction}

\begin{enumerate}[label=D\arabic*), ref=D\arabic*]
    
    \item \label{conditions1_dbg} $k$ must be strictly greater than the maximum length of a double repeat.
    
    \item \label{conditions2_dbg} In each haplotype, every $k + 1$ length substring of that haplotype must be covered by a read. 

\end{enumerate}

\subsubsection{Proof of the sufficiency of Conditions~\ref{conditions1_dbg} and \ref{conditions2_dbg}}

\begin{figure}[hbt!]
    \centering
    \includegraphics[scale=0.15]{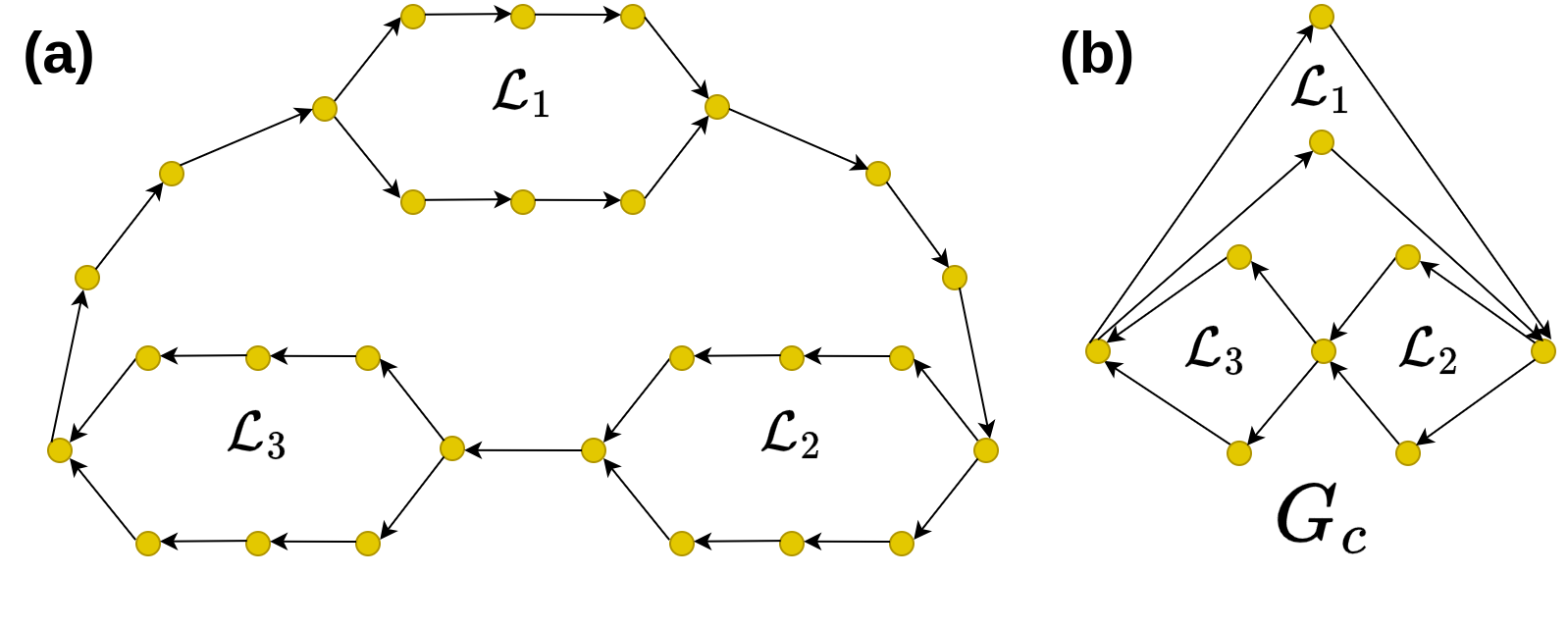}
    \caption{\textbf{(a)}~An example of a de Bruijn graph when $k$ is strictly greater than the length of any double repeat. \textbf{(b)}~The corresponding condensed sequence graph.}
    \label{diploid-debruijn-doublerepeatsbridged}
\end{figure}

\begin{Claim}
    All nodes in the de Bruijn graph have in-degree and out-degree at most $2$.
\end{Claim}
    
\begin{proof}
    We first assume that at least $k$ characters separate any two adjacent heterozygous loci and later extend the proof for the general case. Since Condition \ref{conditions1_dbg} ensures that all the double repeats are bridged, this implies the only nodes having in-degree (out-degree) $2$ are the ones that correspond to the $k$-mer starting (ending) right after (before) a heterozygous locus. From Condition \ref{conditions2_dbg}, the remaining degrees will exactly be $1$ for the graph to remain strongly connected. Figure $\ref{diploid-debruijn-doublerepeatsbridged}$(a) shows an example. Suppose less than $k$ characters separate a pair of adjacent loci. In that case, $k$-mers covering those loci positions will be a part of a single bubble; hence, the overall bubble-graph structure of the de Bruijn graph will not be disturbed. 
\end{proof}

\begin{Claim}
    The condensed sequence graph $G_c$ is Eulerian, and the Eulerian tour of the same gives the concatenation of the two haplotypes up to switch errors.
\end{Claim}

\begin{proof}
    If Conditions \ref{conditions1_dbg} and \ref{conditions2_dbg} hold, the in-degree and out-degree of every node in $G_c$ is equal, making the graph Eulerian (refer Figure $\ref{diploid-debruijn-doublerepeatsbridged}$(b) for reference). The second part of the proof can be deduced from Claim \ref{claim_removeoverlap}.
\end{proof}

% \begin{Remark}
%     Necessary conditions are similar to sufficient conditions except for one corner case, where the value of $k$ can be relaxed. We analyze this case in Section \ref{dbg-specialcase} of the appendix.
% \end{Remark}  

\subsubsection{Necessary conditions for correct reconstruction}

\begin{enumerate}[label=N\arabic*), ref=N\arabic*]
    
    \item \label{condition1_dbg} $k$ must be strictly greater than the maximum length of a double repeat, excluding the intra-double repeats in which both the copies share a common heterozygous locus. 
    
    \item \label{condition2_dbg} For each haplotype, every $k + 1$ length substring of that haplotype must be covered by a read.

\end{enumerate}

% The proof of necessity is given in Section \ref{dbg-necessary} of Appendix.
\subsubsection{Proof of the necessity of Conditions~\ref{condition1_dbg} and \ref{condition2_dbg}}
\label{dbg-necessary}

\begin{Claim}
    Condition \ref{condition1_dbg} is necessary for unique reconstruction by Algorithm \ref{dbg-algorithm}. 
\end{Claim}

\begin{proof}

\begin{figure}[htbp]
    \centering
    \includegraphics[scale=0.17]{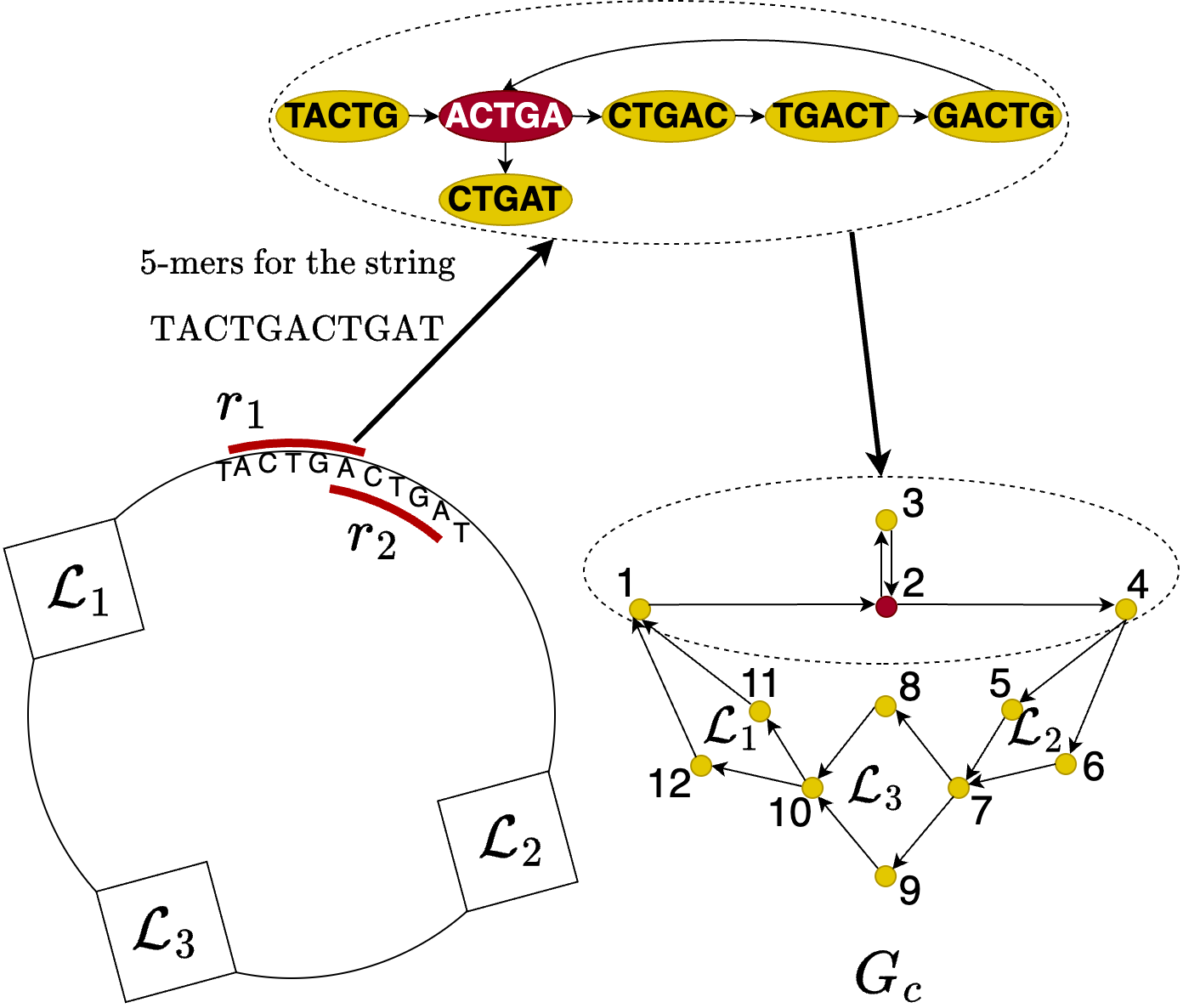}
    \caption{Figure shows one possible arrangement of double repeats on the bubble graph where the two copies $(r_1, r_2)$ belong to a homozygous region and overlap with each other. The two copies are highlighted in red. The condensed sequence graph ($G_c$) is also shown on the right. $5$-mers for a string in the homozygous region are shown along with the respective edges in the top bubble to explain the topology of condensed graph $G_c$.} The above example where vertices $1$ and $4$ have odd degrees confirms that when $k$ is equal to the length of the double repeat, $G_c$ is not guaranteed to be Eulerian.
    \label{diploid-debruijn-case1}
\end{figure}

Similar to the proof of Claim \ref{claim2_greedy}, we analyze all the arrangements of double repeats, excluding intra-double repeats, with both copies having at least one heterozygous locus in common. We look at one case in some detail in Figure \ref{diploid-debruijn-case1}. Four other cases are shown in  Figure \ref{diploid-debruijn} in the Appendix. Three other cases are not shown for which the condensed sequence graphs have topologies similar to those shown in the figures. For $k$ equal to the length of the double repeat, all the cases fail to guarantee unique reconstruction up to switch errors. % The exceptional case is analyzed in Section \ref{dbg-specialcase} of this appendix.
\end{proof}

\begin{Claim}
    Condition \ref{condition2_dbg} is necessary for unique reconstruction by Algorithm \ref{dbg-algorithm}. 
\end{Claim}

\begin{proof}
Similar to the haploid case \cite{bresler2013optimal}, this condition ensures that the de Bruijn graph is strongly connected, i.e., every vertex can be visited from every other vertex. Strong connectivity ensures that the condensed sequence graph contains an edge-covering closed walk.     
\end{proof}

The necessary condition \ref{condition1_dbg} specifically excludes arrangements of intra-double repeats with both copies having at least one heterozygous locus in common. This is because to guarantee the reconstruction of regions in the genome containing such repeats, a weaker condition on $k$ is necessary, as we explain in the next subsection.

\subsubsection{Special case}
\label{dbg-specialcase}

Here we analyze the special case of intra-double repeat with both copies $r_1, r_2$ having at least one heterozygous locus in common. Let $o$ denote the length of overlap between the repeats $r_1$ and $r_2$, where $r_1$ occurs before $r_2$ when arranged in clockwise order of the starting positions. Suppose $\textsc{union}(r_1, r_2)$ cover the loci $\mathcal{L}_1, \mathcal{L}_2, \ldots, \mathcal{L}_m$, then $d$ is the distance from $\mathcal{L}_m$ to the end of $r_2$ (both inclusive).
% , we define the following notations and show them in Figure \ref{diploid-debruijn-exception-notation}.

% \begin{itemize}
%     \item $\mathcal{L}_i:$ $i^{th}$ heterozygous locus with bases $X_i, Y_i$ on opposite haplotypes. 
    
%     \item $g_i:$ Gap between (including the ends) heterozygous loci $\mathcal{L}_i$ and $\mathcal{L}_{i + 1}$.
    
%     \item $o:$ Length of overlap between the repeats $r_1$ and $r_2$, where $r_1$ occurs before $r_2$ when arranged in clockwise order of the starting positions. 
    
%     \item $d:$ Suppose $\textsc{union}(r_1, r_2)$ cover the loci $\mathcal{L}_1, \mathcal{L}_2, \ldots, \mathcal{L}_m$, then $d$ is the distance from $\mathcal{L}_m$ to the end of $r_2$.
% \end{itemize}
    
To guarantee unique reconstruction up to switch errors, we must have $k \geq \max((\max_{i=1}^{m-1} g_i) + 1, d)$. In the Appendix, we prove this result through case analysis for $m = 1$ in Figure \ref{diploid-debruijn-exception}. The analysis can be extended for $m > 1$.

\subsection{Overlap graph}
% \input{sectionfiles/algorithms/overlap_sufficient}

% \subsubsection{Sufficient conditions for unique reconstruction without switch errors}

% \begin{figure}[hbt!]
%     \centering
%     \includegraphics[scale=0.15]{images/DiploidWithPhasing-OLC.png}
%     \caption{Figure (A) represents the case when the two repeat copies cover distinct heterozygous loci, and Figure (B) represents when the two copies cover the same loci but belong to different haplotypes. Double repeats are not bridged for both cases, resulting in a possible reconstruction with switch error, as shown in the sparse read overlap graphs.}
%     \label{DiploidWithPhasing-OLC}
% \end{figure}

% \begin{enumerate}
%     \item For repeats of the type mentioned in Figure \ref{DiploidWithPhasing-OLC}: Double repeats must be bridged.
% \end{enumerate}
\subsubsection{Algorithm}
%
%\NK{This description needs to be improved. Firstly, it says that the algorithm works by constructing a transitively reduced \cite{myers2005fragment} all-vs-all read overlap graph; but it looks like such a reduced overlap graph is in fact the input to the algorithm. Then, there is a mismatch in the voice: the first sentence says that the algorithm does something, while the second sentence sarts with ``Next, we [do something]''. Also, I think you should define what a ``transitively reduced all-vs-all read overlap graph'' is. Please re-write this paragraph more carefully.} 
A read overlap graph is constructed by adding directed edges between a pair of nodes (reads) if they share a suffix-prefix overlap, and the weight of the edge is considered to be the length of the maximum overlap. In the overlap graph, if $x$ overlaps $y$, $y$ overlaps $z$, $x$ overlaps $z$, and $\textsc{union}(\textsc{union}(x,y), z) = \textsc{union}(x,z)$, then the edge from $x$ to $z$ is said to be \emph{transitively inferable}. Such edges are redundant and are therefore excluded from the graph \cite{myers2005fragment}. Overlap graph $G'$ obtained after removing all the transitively inferable edges is fed as an input to the assembly algorithm. Appealing to parsimony, one way to formulate the assembly problem is to find the shortest node-covering closed walk in graph $G'$. Finally, the string spelled by this walk is processed similarly to Algorithm \ref{dbg-algorithm} to report the reconstructed haplotypes.

% \begin{figure}[hbt!]
\begin{figure*}
    \centering
    \includegraphics[scale=0.22]{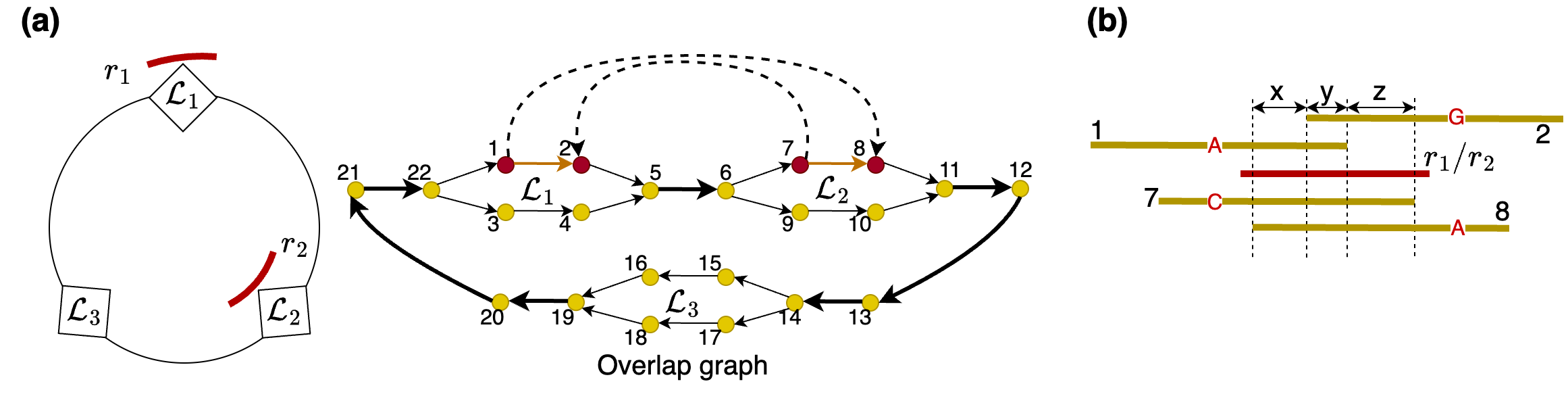}
    \caption{Figure \textbf{(a)} shows the case when there is a double repeat with both copies covering a heterozygous locus. Also, a read-overlap graph is shown without transitively inferable edges. Solid edges are the correct edges, while the dashed edges are the incorrect ones. Figure \textbf{(b)} shows the reads $1, 2, 7$ and $8$ which overlap with the repeat $(r_1, r_2)$. Here $(1 \rightarrow 2, 7 \rightarrow 8)$ is the correct overlap, but the combined length of overlap is exactly equal to $(1 \rightarrow 8, 7 \rightarrow 2)$, i.e., $x + 2y + z$. Hence, for any given set of reads, the given arrangement of repeats will have at least two possible reconstructions.}
    \label{doublerepeatsbridged}
\end{figure*}

% \subsubsection{Algorithm}
\begin{algorithm}[htpb]
\small
\caption{: Overlap graph-based algorithm for diploid genomes}
\label{overlap-algorithm}
\begin{algorithmic}[1]
\State \textbf{Input:} Overlap graph $G'$
\State \textbf{Output:} Reconstructed haplotypes $(\mathcal{G}_0, \mathcal{G}_1)$
\State $A :$ Array where $A[i]$ stores the string spelled by the vertex indexed $i$ in $G'$.
\State \label{node-covering-walk}$\textsc{WALK} :$ List of vertices in the order visited by the shortest node covering closed walk of $G'$, where all vertices are distinct (the starting vertex is added only once).

\State $l \leftarrow \lvert \textsc{WALK} \rvert$
\State $S \leftarrow A[\textsc{WALK}[0]]$
\For{$i = 1 \text{ to } l - 1$}
    \State $o \leftarrow \textsc{overlap}(\textsc{WALK}[i - 1], \textsc{WALK}[i])$
    \State $S \leftarrow S + A[\textsc{WALK}[i]][o :]$
\EndFor
\State $o \leftarrow \textsc{overlap}(\textsc{WALK}[l - 1], \textsc{WALK}[0])$
\State $S \leftarrow S[: \lvert S \rvert - o]$
\State return $(S[: \lvert S \rvert / 2], S[\lvert S \rvert / 2 :])$
\end{algorithmic}
\end{algorithm}

\subsubsection{Necessary conditions for unique reconstruction up to switch errors}
\label{sufficient_conditions_overlap}
    \begin{enumerate}[label=U\arabic*), ref=U\arabic*]
        \item \label{condition1-overlap} The information-theoretic necessary conditions stated in Section \ref{anchor_uptoerrors} must hold.
        \item \label{condition2-overlap} Triple repeats must be bridged.
        \item \label{condition3-overlap} Double repeats, excluding the intra-double repeats in which both the copies share a common heterozygous locus, must be bridged.
    \end{enumerate}

The necessity of condition \ref{condition2-overlap} can be inferred from the haploid case \cite{bresler2013optimal}.

\subsubsection{Proof of the necessity of Condition~\ref{condition3-overlap}}
\label{overlapnecessity}

\begin{Claim}
    Condition \ref{condition3-overlap} is necessary for unique reconstruction up to switch errors by algorithm \ref{overlap-algorithm}.
\end{Claim}

\begin{proof}
Similar to the proof of Claim \ref{claim2_greedy}, we analyze all the arrangements of double repeats, excluding intra-double repeats, with both copies having at least one heterozygous locus in common. We look at one case in some detail in Figure \ref{doublerepeatsbridged}. Five other cases are shown in  Appendix Figure \ref{OLC-final-vertical}. Two other cases are not shown for which the read-overlap graphs have topologies similar to those shown in the figures.

Figure \ref{doublerepeatsbridged}(a) shows a read-overlap graph after the removal of transitively inferable edges. If the double repeats are not bridged, there exists an alternate reconstruction ($1 \rightarrow 8 \rightarrow 11 \rightarrow 12 \rightarrow 13 \rightarrow 14 \rightarrow 15 \rightarrow 16 \rightarrow 19 \rightarrow 20 \rightarrow 21 \rightarrow 22 \rightarrow 3 \rightarrow 4 \rightarrow 5 \rightarrow 6 \rightarrow 7 \rightarrow 2 \rightarrow 5 \rightarrow 6 \rightarrow 9 \rightarrow 10 \rightarrow 11 \rightarrow 12 \rightarrow 13 \rightarrow 14 \rightarrow 17 \rightarrow 18 \rightarrow 19 \rightarrow 20 \rightarrow 21 \rightarrow 22 \rightarrow 1$) which is not the same as the ground truth, even after allowing for switch errors. Also, the alternate reconstruction has exactly the same length as that of the ground truth, which is evident from Figure \ref{doublerepeatsbridged}(b). 
\end{proof}

\section{Experiments}\label{sec:experiments}
\label{experiments}
%\subsection{Data preparation and results}
\label{plotdata}

% Maternal and paternal copy of the human chromosome $19$ is taken from the HG002 reference genome (https://github.com/marbl/HG002). Next, variant calling is done using dipcall (https://github.com/lh3/dipcall) by taking the maternal copy as a reference, and the SNP positions are identified. We modify the maternal copy at these SNP positions to get the paternal copy for finding the repeat statistics listed in table \ref{repeatstats}. 

We apply our theoretical analysis to compute the coverage depth and read length requirements for the assembly of human chromosome $19$. For this experiment, we used an already assembled sequence of the maternal copy of chromosome $19$ in the HG002 genome. This sequence was assembled by the Telomere-to-telomere consortium  \cite{marblHG002}. %\hl{The publicly-available variant calls in HG002 genome are only available in the high-confidence regions, The high quality variant calls are not readily available for human genomes (e.g., HG002) in several challenging (difficult-to-map) regions of the genome} \cite{wagner2022benchmarking}.
%We create two instances of a diploid chromosome by substitution events at every position independently with probability $\Pr = 0.001$ and $\Pr = 0.0001$ to obtain the paternal copy. 
We simulated heterozygous substitution events to obtain the paternal sequence. We opted to simulate these variants rather than using real variation data because the set of high-quality variant calls in HG002 genome released by the Genome in a Bottle consortium are only available in a fraction of the genome (high confidence intervals).

In the simulation, we selected each position as a heterozygous locus independently with some probability. We used probability values $\Pr = 0.001$ and $\Pr = 0.0001$, respectively, to obtain two diploid chromosomes with different heterozygosity rates. At each heterozygous locus, we chose the character in the paternal sequence uniformly randomly from the three available choices. 
%\NK{It is not clear what this is intended to mean. Please give a precise description of the model used fro simulating SNPs.}
Next, we computed the repeat statistics, shown in Table \ref{repeatstats}.
%\subsection{Results}
Using the repeat statistics and the conditions proven in Sections \ref{sec:it}-\ref{sec:algo}, we make the following inferences:
\begin{itemize}
    \item Lower bound on the read length required for unique reconstruction up to switch errors:
    \begin{enumerate}
        \item For the diploid chromosome generated using $\Pr = 0.001$, the length is $\max(8764, 3013, 3432, 9317, 2784) + 2 = 9319$.
        \item For the diploid chromosome generated using $\Pr = 0.0001$, the length is $\max(8764, 6881, 8764, 9317, 5184) + 2 = 9319$.
        % \item For $\Pr = 0.0002: \max(8764, 4848, 7472, 9317, 4167) + 2 = 9319$.
        % \item For Chromosome $1: \max(18551, 3119, 4091, 8983, 2789) + 2 = 18553$.
    \end{enumerate}
    \item The minimum read length for the greedy algorithm to reconstruct the genome up to switch errors:
    \begin{enumerate}
        \item For the diploid chromosome generated using $\Pr = 0.001$, the length is  $\max(\left\lceil \frac{12748 + 3}{2} \right\rceil, 16812, 9319) = 16812$.
        \item For the diploid chromosome generated using $\Pr = 0.0001$, the length is $\max(\left\lceil \frac{78519 + 3}{2} \right\rceil, 33145, 9319) = 39261$.
        % \item For $\Pr = 0.0002: \max(\left\lceil \frac{55374 + 3}{2} \right\rceil, 16812, 9319) = 27689$.
        % \item For Chromosome $1: \max(\left\lceil \frac{12361 + 3}{2} \right\rceil, 19823, 18553) = 19823$.
    \end{enumerate} 
    \item The minimum value of $k$ for the de Bruijn graph algorithm to reconstruct the genome up to switch errors:
    \begin{enumerate}
        \item For the diploid chromosome generated using $\Pr = 0.001$, the value is 
        $\max(16810 + 1, \ 9319) = 16811$.
        \item For the diploid chromosome generated using $\Pr = 0.0001$, the value is 
        $\max(16810 + 1, \ 9319) = 16811$.
        % \item For $\Pr = 0.0002: \max(16810 + 1, 9319) = 16811$.
        % \item For Chromosome $1: \max(19821 + 1, 18553) = 19822$.
    \end{enumerate}
\end{itemize}

\begin{table*}[hbt!]
  \centering
  % \begin{tabular}{|p{6.5cm}|c|}
  % \fontsize{9pt}{9pt}
    \begin{tabular}{|c|c|c|c|}
    \hline
    \textbf{Parameter} & \textbf{Value ($\Pr = 0.001$)} & \textbf{Value ($\Pr = 0.0001$)}\\%& \textbf{Value ($\Pr = 0.0002$)} & \textbf{Value (Chromosome $1$)} \\
    \hline
    Length of each haplotype & $61,317,360$ & $61,317,360$\\%& $61,317,360$ & $244022132$\\
    \hline
    Maximum gap between two adjacent heterozygous loci & $12,748$ & $78,519$\\%& $55,374$ & $12361$\\
    \hline
    Maximum length of a double repeat & $16,810$ & $16,810$\\%& $16,810$ & $19821$\\
    \hline
    Minimum read length required for all double repeats to be well-bridged & $16,812$ & $33,145$\\%& $16812$ & $19823$\\
    \hline
    Maximum length of an interleaved repeat in the maternal haplotype & $8,764$ & $8,764$\\%& $8,764$ & $18551$\\
    \hline
    Maximum length of an interleaved repeat in the paternal haplotype & $3,013$ & $6,881$\\%& $4,848$ & $3119$\\
    \hline
    Maximum length of a repeat of the type mentioned in \ref{information_theoretic_intercondition} & $3,432$ & $8,764$\\%& $7,472$ & $4091$\\
    \hline
    Maximum length of a triple repeat in the maternal haplotype & $9,317$ & $9,317$\\%& $9,317$ & $8983$\\
    \hline
    Maximum length of a triple repeat in the paternal haplotype & $2,784$ & $5,184$\\%& $4,167$ & $2789$\\
    \hline
  \end{tabular}
  \caption{Repeat statistics for human diploid chromosome $19$. We show the statistics for both diploid chromosomes simulated with different heterozygosity rates. Our code for computing repeat statistics is available at \href{https://github.com/at-cg/DiploidGenomeAssembly}{https://github.com/at-cg/DiploidGenomeAssembly}.}
  \label{repeatstats}
\end{table*}

\begin{figure}[htpb]
% \begin{figure*}
    \centering
    \includegraphics[scale=0.33]{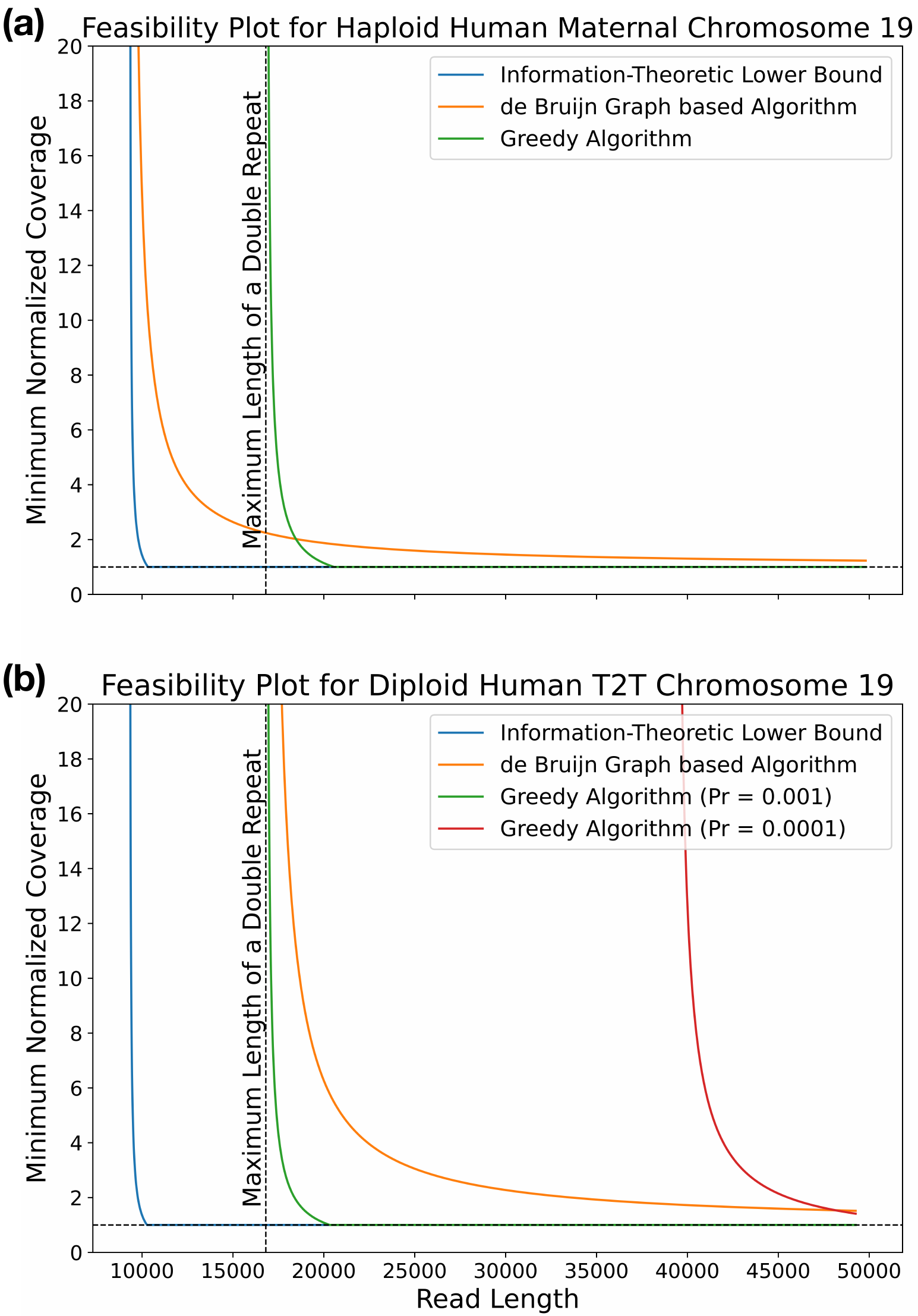}
    \caption{Feasibility plot for the reconstruction of (a) haploid and (b) diploid human chromosome $19$ up to switch errors with probability at least $0.99$.}
    \label{feasibilityplot}
\end{figure}

In Appendix Section \ref{coverage_analysis}, we also derive conditions on the minimum coverage depth and read length required for a unique reconstruction up to switch errors with probability at least $1 - \epsilon$. Our analysis is partly inspired by the coverage analysis of Bresler \emph{et al.} \cite{bresler2013optimal} for haploid genome reconstruction.
%Using Eq. \eqref{itplot}, \eqref{greedyplot}, \eqref{dbgplot} derived in Appendix, we get the feasibility plot shown in Figure \ref{feasibilityplot}. We validate that the values which satisfy Eq. \eqref{greedyplot} also satisfy Eq. \eqref{greedy_a}.
Using the equations, we obtained the feasibility plot shown in Figure \ref{feasibilityplot}(b). We also show the feasibility plot for haploid genome construction in Figure \ref{feasibilityplot}(a) by using the equations of Bresler \emph{et al.} \cite{bresler2013optimal}. Having both plots makes it convenient to understand the difference between information-theoretic conditions for haploid and diploid genome construction. 

Comparing Figure \ref{feasibilityplot}(a) with Figure \ref{feasibilityplot}(b), we observe that the orange curve associated with the de Bruijn graph-based algorithm shifts right as we increase ploidy. 
%This is because double repeats form the bottleneck for the diploid genome reconstruction algorithms. 
The conditions of the de Bruijn graph-based algorithm for haploid genome reconstruction are with respect to interleaved and triple repeats \cite{bresler2013optimal}, whereas the conditions that we derived involve double repeats of the target diploid genome. By definition, bridging of all double repeats ensures that all interleaved and triple repeats are bridged, but the opposite does not hold.

Next, we look at Figure \ref{feasibilityplot}(b) in more detail. On our dataset generated using $\Pr = 0.001$, the greedy algorithm performs better than the de Bruijn graph-based algorithm in terms of minimum coverage depth requirements because (i) the de Bruijn graph-based algorithm has a strict requirement that every $k + 1$ length substring of a haplotype must be covered by a read, and (ii) the maximum gap between adjacent loci is small enough to not affect the coverage requirement of greedy algorithm. However, the coverage requirements of the greedy algorithm increase on our second dataset generated using $\Pr = 0.0001$. The increase in the length of the maximum gap between two adjacent heterozygous loci leads to an increase in the coverage requirements of the greedy algorithm to satisfy condition \ref{condition2.1_greedy}. This condition asks for a pairwise overlap between reads that cover adjacent heterozygous loci. 

Lastly, we note that all the above observations also apply to the assembly of other sequencing datasets where the target genome has similar heterozygosity rate and repeat length statistics.

\section{Conclusions and Future Work}
In this paper, we present the first rigorous information-theoretic analysis for the problem of diploid genome assembly. Diploid genome assembly is a fundamental problem in computational biology; for example, it is used to study an individual's genome accurately without reference bias and to build human pangenome references. Compared to haploid genomes, the information-theoretic analysis for diploid genomes is more challenging due to the presence of heterozygous variation and the increase in the number of classes and arrangements of repeats. 

The proposed information-theoretic conditions in Section \ref{sec:it} do not deviate significantly from the already known conditions for haploid genome reconstruction. We do not expect our Condition \ref{information_theoretic_intercondition} to influence the read length and coverage requirements on real genome sequences. This is also evident from our experiments using human chromosome 19. 

While analysing the standard assembly algorithms in Section \ref{sec:algo}, we showed that their information requirement for the correct reconstruction of a diploid genome is significantly higher compared to the known conditions for a haploid genome. The performance of these algorithms is limited by the maximum length of the double repeat. Meeting these read length and coverage requirements with error-free reads is currently infeasible in practice, which justifies the need for heuristics for generating complete telomere-to-telomere genome assemblies.

We highlight a few possible directions for future work. First, this paper does not provide information-theoretic sufficient conditions for overlap graph-based algorithm as we found their derivation challenging. Second, the wide gap between the lower bound and the information-theoretic conditions of the known assembly algorithms suggests that there may be alternative information-optimal assembly algorithms for diploid genome assembly. Also, accounting for the noise in sequencing data and considering more complex variations creates more exciting lines of work in this area.

Another interesting future direction is to determine the information requirements for reconstructing polyploid genomes. Many of our definitions, such as those for different repeat types, remain applicable to polyploid genomes. Additionally, our information-theoretic necessary conditions still hold because polyploid genome reconstruction is a harder problem than its diploid counterpart. However, the assembly algorithms considered in this paper would need to be adapted for polyploid genome reconstruction before analysis. Notably, in a polyploid genome (ploidy $>2$), two or more haplotypes can share the same allele at a heterozygous locus, which may need to be explicitly considered in the algorithms and their analysis.

\section*{Acknowledgement}
We thank the Telomere-to-Telomere (T2T) consortium for making their sequencing datasets and genome assemblies publicly available. This research is partly supported by the funding from the DBT/Wellcome Trust India Alliance (IA/I/23/2/506979). We also thank the Ministry of Education, India, for the financial support through the Prime Minister's Research Fellowship (PMRF). We used computing resources provided by the National Energy Research Scientific Computing Center (NERSC), USA.

% \input{sectionfiles/guidelines}

% \input{bibliography}
% \input{main.bbl}

% \input{sectionfiles/authors}

\appendix %\section{Appendix}

\subsection{Coverage Analysis} \label{coverage_analysis}
Define diploid genome length $G \coloneqq 2 \lvert \mathcal{H} \rvert$. 
If the read length is equal to $L$ and the starting location of $N$ reads follow a Poisson process with rate $\lambda = \frac{N}{G}$, then the probability that a repeat of length $i$ is unbridged (denoted as $p_i^{\text{unbridged}}) $ is $e^{-2\lambda(L - i - 1)}$. 
The gap, i.e., the count of characters between the pair of adjacent heterozygous loci $(\mathcal{L}_i, \mathcal{L}_{i + 1})$ (both exclusive) is denoted as $g_i$.

\subsubsection{Information-theoretic necessary conditions} \label{it_coverage}

First, we calculate the probability of making an error in the genome reconstruction due to intra-repeats. Let $b_{mn}$ denote the number of intra-interleaved repeats of length $m, n$ (where $m < n$) and $c_p$ denote the number of intra-triple repeats of length $p$ across both the haplotypes. 

As presented in \cite{bresler2013optimal}, probability of error $P_{\text{err}}^1$ can be written as follows:

\[
  P_{\text{err}}^1 \approx \frac{1}{2} \max \left\{
  \begin{array}{ll}
    \sum_{\substack{m < L - 1 \\ n \geq L - 1}} b_{mn}e^{-2\lambda(L - m - 1)} \\ \\
    \sum_{m, n < L - 1} b_{mn}e^{-2\lambda(L - m - 1)}e^{-2\lambda(L - n - 1)} \\ \\
    \sum_{p} c_{p}e^{-3\lambda(L - p - 1)}
  \end{array}
  \right.
\]

For desired reconstruction probability, $P_{\text{err}}^1$ must be strictly less than $\epsilon$, thereby getting a lower bound on $L$ as a function of $L$ and $\lambda$:

\[
  L^{(1)} > \max \left\{
  \begin{array}{ll}
    \frac{1}{2\lambda}\ln(\frac{\gamma_1}{2\epsilon}), \text{where } \gamma_1 = \sum_{\substack{m < L - 1 \\ n \geq L - 1}}b_{mn}e^{2\lambda(m + 1)} \\ \\
    \frac{1}{4\lambda}\ln(\frac{\gamma_2}{2\epsilon}), \text{where } \gamma_2 = \sum_{m, n < L - 1}b_{mn}e^{2\lambda(m + n + 2)} \\ \\
    \frac{1}{3\lambda}\ln(\frac{\gamma_3}{2\epsilon}), \text{where } \gamma_3 = \sum_{p}c_{p}e^{3\lambda(p + 1)}
  \end{array}
  \right.
\]

Next, we calculate the error probability due to inter-double repeats as mentioned in Condition \ref{information_theoretic_intercondition} in Section~\ref{anchor_uptoerrors}.
% We only analyze the same for $m = 1$, since the scenario $m > 1$ is unlikely to happen in real genomes. 
Let $d_{xy}$ denote the number of pairs where one double repeat is of length $x$, and the other double repeat is of length $y$ (where $x \leq y$) and $\lvert s_1 \rvert = \lvert s_2 \rvert > 0$. The analysis remains similar to that of intra-interleaved repeats. Again for the error to be strictly less than $\epsilon$, we get a lower bound on $L$: 

% \[
%   P_{\text{err}}^2 \approx \frac{1}{2} \cdot \max \left\{
%   \begin{array}{ll}
%     \sum_{\substack{x < L - 1 \\ y \geq L - 1}} d_{xy}e^{-2\lambda(L - x - 1)} \\ \\
%     \sum_{x, y < L - 1} b_{xy}e^{-2\lambda(L - x - 1)}e^{-2\lambda(L - y - 1)}
%   \end{array}
%   \right.
% \]

\[
  L^{(2)} > \max \left\{
  \begin{array}{ll}
    \frac{1}{2\lambda}\ln(\frac{\gamma_4}{2\epsilon}), \text{where } \gamma_4 = \sum_{\substack{x < L - 1 \\ y \geq L - 1}}d_{xy}e^{2\lambda(x + 1)} \\ \\
    \frac{1}{4\lambda}\ln(\frac{\gamma_5}{2\epsilon}), \text{where } \gamma_5 = \sum_{x, y < L - 1}d_{xy}e^{2\lambda(x + y + 2)} 
  \end{array}
  \right.
\]

Finally, we want read length $L \geq \max(L^{(1)}, L^{(2)})$. Also, it is clear from the data in Section \ref{experiments} that the value of $\gamma_i, \forall i \in [5]$ will be dominated by the longest repeat; therefore, other repeats can be neglected. Making this simplification, we get $N$ as a function of $L$ as follows:

% \[
\begin{equation} \label{itplot}
    \resizebox{\columnwidth}{!}{$ 
    N^{(1)} > \max \left\{
  \begin{array}{ll}
    \frac{G}{2 (L - m^* - 1)}\ln\left(\frac{b_{m^*n^*}}{2\epsilon}\right), \\ \text{where } m^* = \max\{m: b_{mn} > 0, m < L - 1, n \geq L - 1\}, \text{and } \\ b_{m^*n^*} = \sum\{b_{mn}: m = m^*, b_{mn} > 0, m < L - 1, n \geq L - 1\} \\ \\
    \frac{G}{4 \left(L - \mu^* - 1\right)}\ln\left(\frac{b^*}{2\epsilon}\right), \\ \text{where } \mu^* = \max\bigl\{\frac12(m + n): b_{mn} > 0, m, n < L - 1\bigr\}, \text{and } \\ b^* = \sum\bigl\{b_{mn}: \frac12(m + n) = \mu^*, b_{mn} > 0, m, n < L - 1\bigr\} \\ \\
    \frac{G}{3 (L - p^* - 1)}\ln\left(\frac{c_{p^*}}{2\epsilon}\right), \text{where } p^* = \max\{p: c_p > 0\} \\ \\
    \frac{G}{2 (L - x^* - 1)}\ln\left(\frac{d_{x^*y^*}}{2\epsilon}\right), \\ \text{where } x^* = \max\{x: d_{xy} > 0, x < L - 1, y \geq L - 1\}, \text{and } \\ d_{x^*y^*} = \sum\{d_{xy}: x = x^*, d_{xy} > 0, x < L - 1, y \geq L - 1\} \\ \\
    \frac{G}{4 \left(L - \xi^* - 1\right)}\ln\left(\frac{d^*}{2\epsilon}\right), \\ \text{where } \xi^* = \max\bigl\{\frac12(x + y): d_{xy} > 0, x, y < L - 1\bigr\}, \text{and } \\ d^* = \sum\bigl\{d_{xy}: \frac12(x + y) = \xi^*, d_{xy} > 0, x, y < L - 1\bigr\} 
  \end{array}
  \right. 
  $}
\end{equation}
% \]

% Probability of having a switch error: 
%  The probability of not having an error is at least equal to the sum of the probability that exactly one pair of adjacent loci is not covered by a read and the probability that all pairs of adjacent loci are covered by a read.

% \begin{align}
%     &\overline{P_{\text{err}}^6} \geq \sum_{i} ((1 - \frac{L - g_i - 1}{G})^N \cdot \prod_{j \neq i}(1 - e^{-\frac{N}{G}(L - g_j - 1)})) + \prod_{i}(1 - e^{-\frac{N}{G}(L - g_i - 1)}) \nonumber \\ 
%     &\ \ \ \ \ = \prod_{i}(1 - e^{-\lambda(L - g_i - 1)}) \cdot (\sum_{i} \frac{(1 - \frac{L - g_i - 1}{G})^{\lambda G}}{(1 - e^{-\lambda(L - g_i - 1)})} + 1) \nonumber \\ 
%     &\ \ \ \ \ \geq 1 - \epsilon
% \end{align}

\subsubsection{Greedy algorithm}

We first analyze the error probability due to Condition \ref{condition2.1_greedy} in Section~\ref{greedy_withoutphasing}. Consider a pair of adjacent loci $(\mathcal{L}_i, \mathcal{L}_{i + 1})$. We fix the right endpoint of read $x_1^i$ (let it be $j$). The probability of sampling such a read equals $\frac{1}{G}$. Next, we find the probability that read $x_2^i$ has its right endpoint ahead of read $x_1^i$ that equals $\frac{g_i - j + 1}{G}$. For condition \ref{condition2.1_greedy} to be satisfied, the homozygous region $[\mathcal{L}_i + j + 1, \mathcal{L}_i + g_j]$ must be bridged by at least one read on both haplotypes. The same can be calculated by first fixing the haplotype where this region is not bridged and then multiplying it with $p_{g_i - j}^{\text{unbridged}}$. A factor of $2$ is multiplied to account for the case when the read $x_1^i$ has its right endpoint ahead of read $x_2^i$. Also, an additional term is added for the gaps where a single read can't cover the adjacent loci, i.e., the probability that the region $[\mathcal{L}_i + 1, \mathcal{L}_i + g_i - L + 1]$ is not bridged by at least one read on at least one of the haplotypes. Using union bound, the total probability of error, $P_{\text{err}}^2$, will be bounded by the sum of these probabilities. Thus, $P_{\text{err}}^2$ is upper bounded by:

\begin{align}
\label{greedy_a}
    &2 \cdot \sum_{i} \left(\sum_{j = max(0, g_i - L + 2)}^{min(g_i, L - 1)} \frac{1}{G} \cdot \frac{g_i - j + 1}{G} \cdot \left(2 \cdot e^{-\frac{N - 2}{G}(L - (g_i - j) - 1)}\right)\right) \nonumber \\
    &+  \sum_{i : (L < g_i + 2)}\left(2e^{-\lambda(L - (g_i - L + 1) - 1)}\right) %\nonumber 
\end{align}    

To simplify our computation, we further upper bound equation \ref{greedy_a} by:

\begin{align}
\label{greedy_a1}
    &2 \cdot \sum_{i} (j_e - j_s + 1) \left(\frac{1}{G} \cdot \frac{g_i - j_s + 1}{G} \cdot \left(2 \cdot e^{-\frac{N - 2}{G}(L - (g_i - j_s) - 1)}\right)\right) \nonumber \\
    &+  \sum_{i : (L < g_i + 2)}\left(2e^{-\lambda(L - (g_i - L + 1) - 1)}\right) %\nonumber 
\end{align}   

where $j_s = max(0, g_i - L + 2)$ and $j_e = min(g_i, L - 1)$. 
It is clear that equation \ref{greedy_a1} is strictly decreasing in $N$ for a fixed $L$; therefore, one can use binary search to find the right value of $N$ given a $L$ and hence compute the contribution of condition $\ref{condition2.1_greedy}$ towards the coverage requirement for the greedy algorithm. 
% \begin{equation}
% \label{greedy_a}
% \resizebox{\columnwidth}{!}{$
%     P_{\text{err}}^2 \leq 2 \cdot \sum_{i} (\sum_{j = max(0, g_i - L + 2)}^{min(g_i, L - 1)} \frac{1}{G} \cdot \frac{g_i - j + 1}{G} \cdot (2 \cdot e^{-\frac{N - 2}{G}(L - (g_i - j) - 1)})) \\
%      +  \sum_{i : (L < g_i + 2)}(2e^{-\lambda(L - (g_i - L + 1) - 1)}) 
%     $}
% \end{equation}

% It is clear from the data in section \ref{plotdata} that analyzing the error probability for the maximum gap suffices. Next, the inner summation can be approximated with integration, and the small constants can be neglected. Making these simplifications and requiring the final term to be strictly less than $\epsilon$ gives us:

% \begin{align}
% \label{complicated}
%     &\frac{4}{(\lambda G)^2}(e^{-\lambda L} + e^{-\lambda(L - g_i)}(\lambda(g_i + 1) - 1)) < \epsilon 
% \end{align}    
    
Next, we look at the error probability due to Condition \ref{condition2.2_greedy}. We divide the repeats into three classes and calculate the error probability (equal to the probability of the repeat not being well-bridged) for them separately. 
Let $A_i$ denote the number of repeats of length $i$ such that both the copies cover a heterozygous locus. Let $B_{ij}$ denote the number of repeats of length $i$ such that exactly one copy covers a heterozygous locus, and the other copy belongs to the gap between the pair of heterozygous loci $(\mathcal{L}_{j}, \mathcal{L}_{j + 1})$. Finally, let $C_{ij_1j_2}$ denote the number of repeats of length $i$ such that none of the copies cover a heterozygous locus, and the copies belong to the gap between the pair of heterozygous loci $(\mathcal{L}_{j_1}, \mathcal{L}_{j_1 + 1})$ and $(\mathcal{L}_{j_2}, \mathcal{L}_{j_2 + 1})$ respectively.

Using a similar analysis as in \ref{it_coverage}, we get:

\[
  L^{(3)} > \max \left\{
  \begin{array}{ll}
    \frac{1}{2\lambda}\ln(\frac{\gamma_6}{\epsilon}), \text{where } \gamma_6 = \sum_{i}A_i e^{2\lambda(i + 1)} \\ \\
    \frac{1}{5\lambda}\ln(\frac{\gamma_7}{\epsilon}), \text{where } \gamma_7 = \sum_{ij}B_{ij} e^{\lambda(i + 1 + 2(i + g_j + 2))} \\ \\
    \frac{1}{8\lambda}\ln(\frac{\gamma_8}{\epsilon}), \text{where } \gamma_8 = \sum_{ij_1j_2}C_{ij_1j_2} e^{2\lambda(g_{j_1} + g_{j_2} + 2i + 4)}
  \end{array}
  \right.
\]

Next, considering only the maximum contributing term in the summation for $\gamma_i, \forall i \in [6, 8]$, we get:

% \[
\begin{equation} \label{greedyplot}
\resizebox{\columnwidth}{!}{$
  N^{(2)} > \max \left\{
  \begin{array}{ll}
    \frac{G}{2 (L - i^* - 1)}\ln\left(\frac{A_{i^*}}{\epsilon}\right), \text{where } i^* = \max\{i: A_i > 0\} \\ \\
    \frac{G}{5 \left(L - \alpha^* - 1\right)}\ln\left(\frac{B^*}{\epsilon}\right), \\ \text{where } \alpha^* = \max\{\frac{1}{5}(3i + 2g_j): B_{ij} > 0\}, \text{and } \\ B^* = \sum\{B_{ij}: \frac{1}{5}(3i + 2g_j) = \alpha^*, B_{ij} > 0\} \\ \\
    \frac{G}{8 \left(L - \beta^* - 1\right)}\ln\left(\frac{C^*}{\epsilon}\right), \\ \text{where } \beta^* = \max\{\frac{1}{4}(g_{j_1} + g_{j_2} + 2i): C_{ij_1j_2} > 0\}, \text{and } \\ C^* = \sum\{C_{ij_1j_2}: \frac{1}{4}(g_{j_1} + g_{j_2} + 2i) = \beta^*, C_{ij_1j_2} > 0\}
  \end{array}
  \right.
  $}
\end{equation}
% \]

\subsubsection{de Bruijn graph-based algorithm}

Let $l_{\text{double}}$ denote the maximum length of a double repeat. Using Equation (6) from \cite{bresler2013optimal}, we get

\begin{equation} \label{dbgplot}
\begin{array}{crc}
\bar{c} = \bigl(1 - \frac{k}{L}\bigr)^{-1} = \bigl(1 - \frac{l_{\text{double}} + 1}{L}\bigr)^{-1}
%    \bar{c} = \frac{1}{1 - \frac{k}{L}} = \frac{1}{1 - \frac{l_{\text{double}} + 1}{L}}
\end{array}
\end{equation}

% \subsection{Greedy Algorithm}
% \input{sectionfiles/Appendix_algos/greedy}

% \subsection{de Bruijn Graph-Based Algorithm}
% \input{sectionfiles/Appendix_algos/deBruijn}

% \subsection{Overlap Graph-Based Algorithm}
% \input{sectionfiles/Appendix_algos/overlap}
% \clearpage
\subsection{Supporting Figures}

\begin{figure*}%[hbt!]
    \centering
    \includegraphics[scale=0.14]{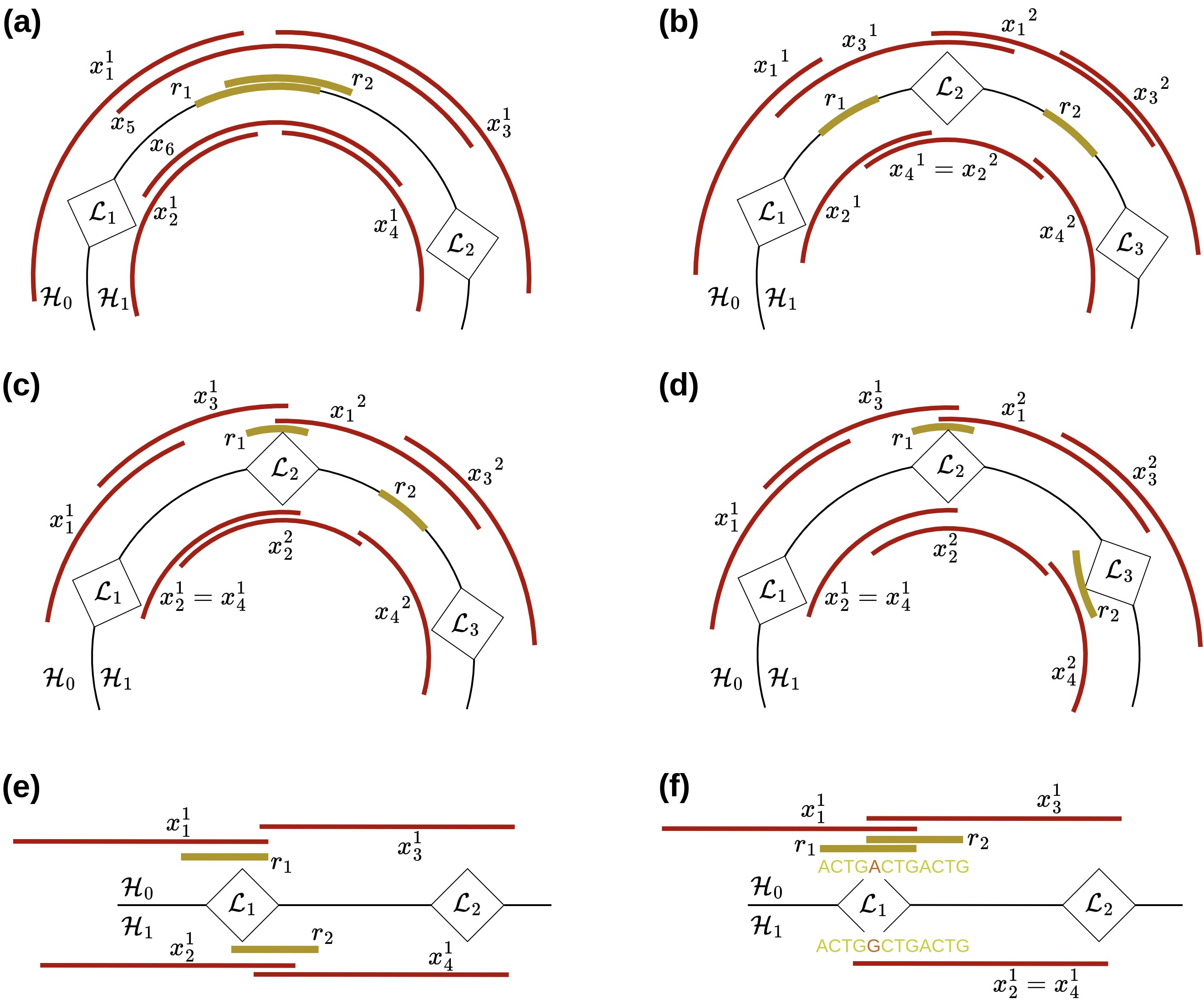}
    \caption{\textbf{Figure shows different possible arrangements of a double repeat; for all cases, the double repeat is not well-bridged.} \textbf{(a)} Both copies of the double repeat $(r_1, r_2)$ lie between the same pair of heterozygous loci and overlap. Note that the reads $x_1^1, x_2^1, x_3^1, x_4^1$ satisfy condition \ref{condition2.1_greedy} because of the presence of repeats. Now, greedy algorithm will merge $x_1^1 \rightarrow x_6 \rightarrow x_5 \rightarrow x_3^1$, $x_2^1 \rightarrow x_4^1$. Although a gap exists between reads $x_2^1$ and $x_4^1$, it is overlooked because of repeats. \textbf{(b)} The double repeat copies belong to the homozygous region between different pairs of heterozygous loci. Here $\textsc{overlap}(x_1^1, x_4^2) > \textsc{overlap}(x_1^1, x_4^1)$ and $\textsc{overlap}(x_1^1, x_4^2) > \textsc{overlap}(x_4^1, x_4^2)$. Greedy algorithm will merge $x_2^1 \rightarrow x_3^1 \rightarrow x_1^2 \rightarrow x_3^2, x_1^1 \rightarrow x_4^2$, thereby leaving read $x_4^1$ unmerged. \textbf{(c)} Exactly one repeat copy of the double repeat covers a heterozygous locus and the two copies do not overlap. Here, $\textsc{overlap}(x_3^1, x_4^2) > \textsc{overlap}(x_3^1, x_1^2)$. Greedy algorithm will merge $x_1^1 \rightarrow x_3^1 \rightarrow x_4^2, x_2^1 \rightarrow x_4^1, x_1^2 \rightarrow x_3^2$. \textbf{(d)} Both the repeat copies cover a different heterozygous locus. Here, $\textsc{overlap}(x_3^1, x_4^2) > \textsc{overlap}(x_3^1, x_1^2)$ and $\textsc{overlap}(x_3^1, x_4^2) > \textsc{overlap}(x_2^2, x_4^2)$. Greedy algorithm will merge $x_1^1 \rightarrow x_3^1 \rightarrow x_4^2, x_4^1 \rightarrow x_2^2, x_1^2 \rightarrow x_3^2$. \textbf{(e)} The repeat copies cover the same heterozygous locus but belong to different haplotypes. Here, the reads $x_1^1, x_4^1$ have the maximum suffix-prefix overlap. The overlap between reads $x_1^1$ and $x_4^1$ is greater because of the presence of repeats than when the repeat is absent. \textbf{(f)} The repeat copies cover the same heterozygous locus and belong to the same haplotype. Like case (e), overlap between reads $x_1^1, x_3^1$ will be greater than when the repeat is absent. Note: Figure (a) can be extended for the case when the repeat copies do not overlap. Similarly, Figure (c) can be extended for the cases when the repeat copies overlap, or the copies do not overlap and have at least one heterozygous locus between the close ends of the copies that is not covered by any of the copies.}
    \label{greedy-extra-condition}
\end{figure*}

\begin{figure*}[hbt!]
    \centering
    \includegraphics[scale=0.2]{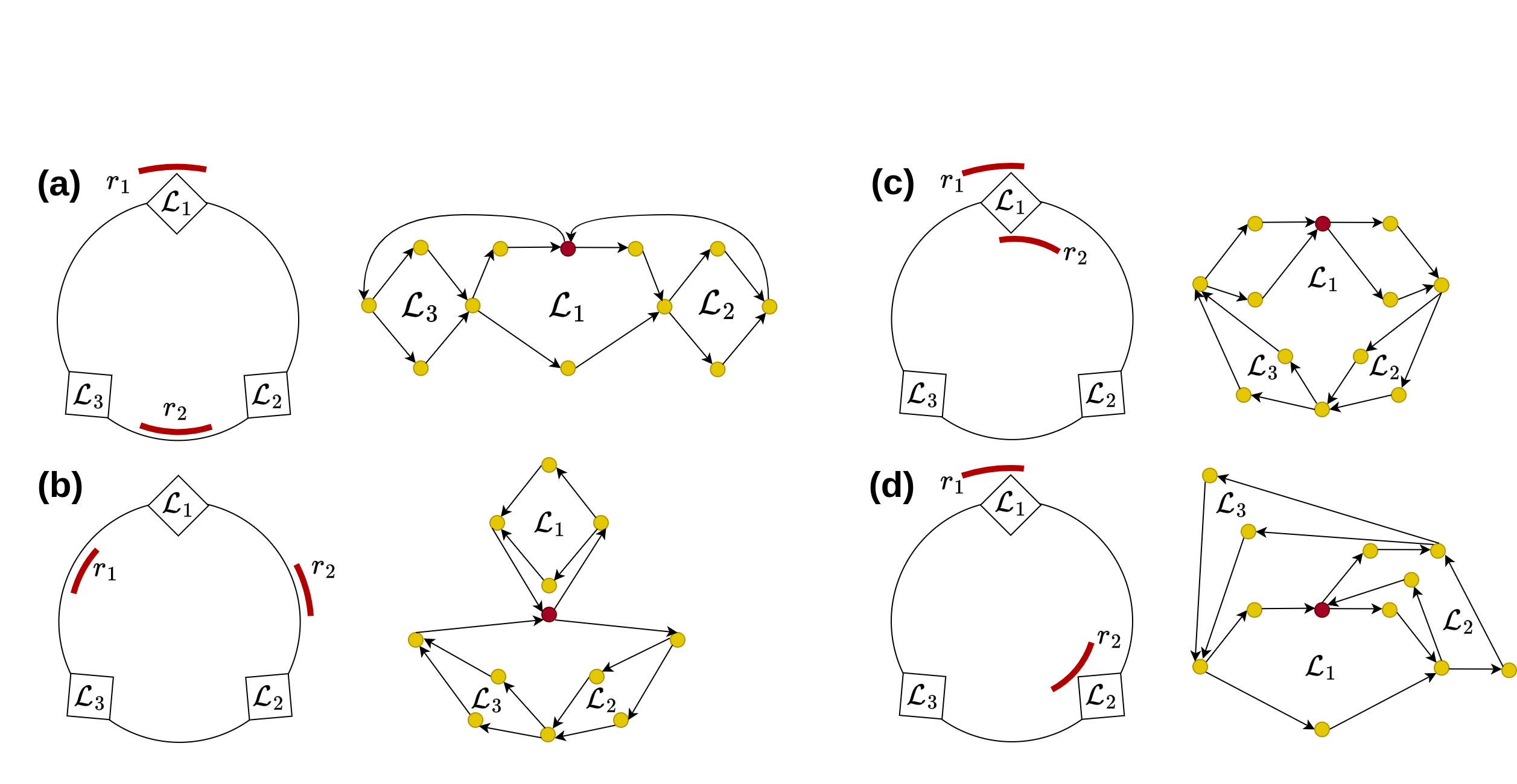}
    \caption{\textbf{Figure shows different possible arrangements (not exhaustive) of double repeats on a diploid genome with three heterozygous loci. Also, for $k$ equal to the length of the double repeat, the condensed sequence graph corresponding to the same is shown.} \textbf{(a)} Exactly one copy covers a heterozygous locus. \textbf{(b)} None of the copies cover a heterozygous locus. \textbf{(c)} Both copies cover the same heterozygous locus but belong to different haplotypes. \textbf{(d)} Both copies cover different heterozygous loci. In (a) and (b), the condensed sequence graph is not Eulerian. In (c) and (d), the condensed sequence graph is Eulerian, but not all Eulerian cycles correspond to unique reconstruction up to switch errors.}
    \label{diploid-debruijn}
\end{figure*}

\begin{figure*}
    \centering
    \includegraphics[scale=0.2]{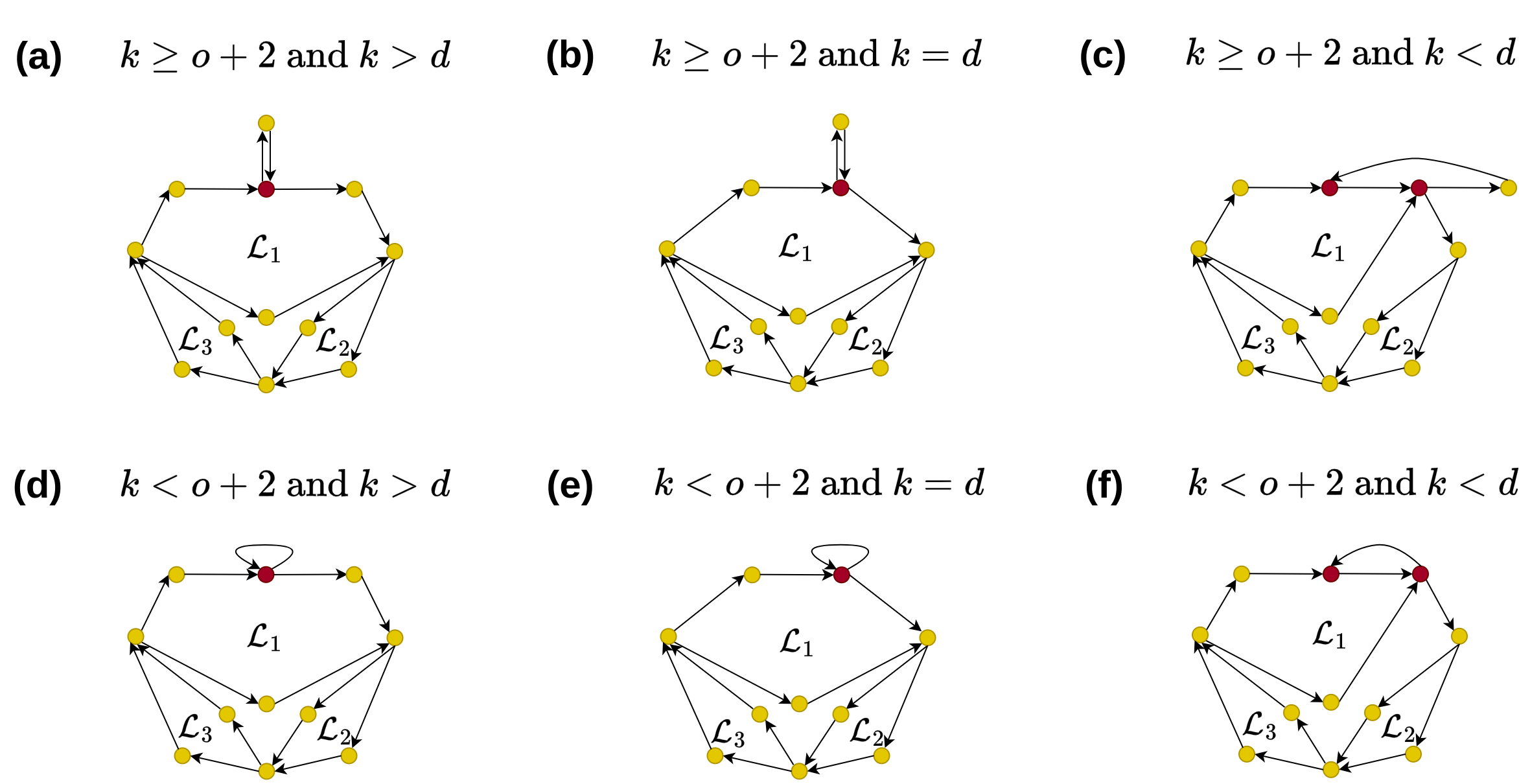}
    \caption{Figure shows the structure of the condensed graph $G_c$ for different values of $k$ when we have an intra-double repeat with both copies having exactly one heterozygous locus ($\mathcal{L}_1$) in common. When \textsc{union}$(r_1, r_2)$ covers exactly one locus, $k \geq d$ is sufficient for unique reconstruction up to switch errors.}
    \label{diploid-debruijn-exception}
\end{figure*}

\begin{figure*}
    \centering
    \includegraphics[scale=0.19]{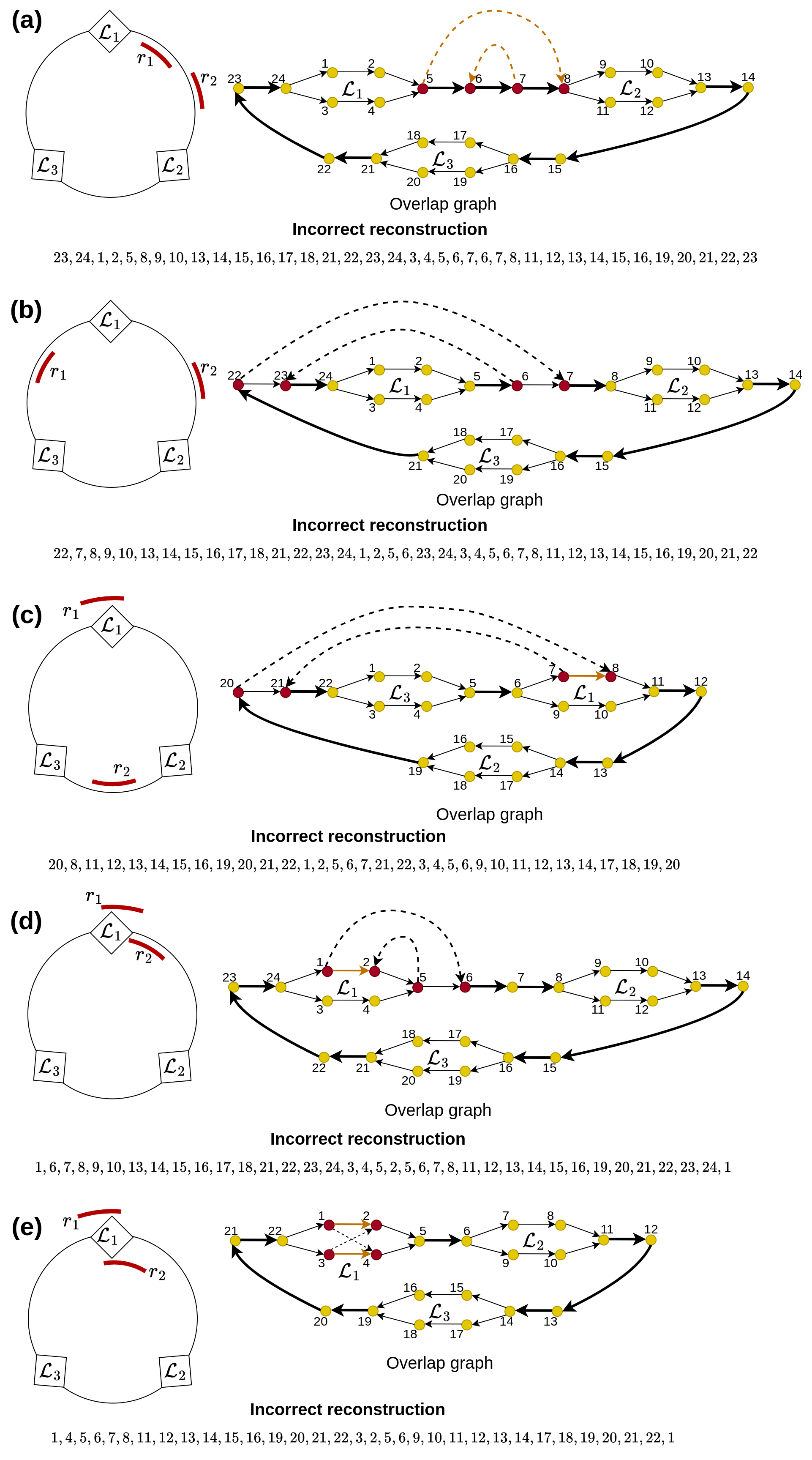}
    \caption{Figure shows different possible arrangements (not exhaustive) of double repeats on a diploid genome with three heterozygous loci. Also, the corresponding read-overlap graph is shown. For all the cases, an alternate reconstruction exists in addition to the ground truth.}
    \label{OLC-final-vertical}
\end{figure*}

\end{document}